\documentstyle[bibnorm,psfig]{lamuphys}
\begin{document}
\title{Fluid Dynamics for Relativistic Nuclear Collisions}

\author{Dirk H.\ Rischke}

\institute{RIKEN-BNL Research Center, Brookhaven National Laboratory,
Upton, NY 11973, USA}

\maketitle

\begin{abstract}
I give an introduction to the basic concepts of
fluid dynamics as applied to the dynamical description of
relativistic nuclear collisions.
\end{abstract}
\section{Introduction and Conclusions}

Modelling the dynamic evolution of nuclear collisions in terms of
fluid dynamics has a long-standing tradition in heavy-ion physics, for
a review see \cite{ris:PRstoecker,ris:PRstrottman,ris:laszlobook}. 
One of the main reasons
is that one essentially does not need more information 
to solve the equations of motion of ideal fluid dynamics than the 
equilibrium equation of state of matter under consideration. 
Once the equation of state is known (and an initial condition is
specified), the equations of motion uniquely determine the dynamics of
the collision. Knowledge about microscopic reaction
processes is not required. This becomes especially important when
one wants to study the transition from hadron to quark and gluon
degrees of freedom, as predicted by lattice simulations of
quantum chromodynamics (QCD) \cite{ris:laermann}. The complicated
deconfinement or hadronization processes need not be known in
microscopic detail, all that is necessary is the thermodynamic
equation of state as computed by e.g.\ lattice QCD.
This fact has renewed interest in fluid dynamics to study the
effects of the deconfinement and chiral symmetry restoration transition
on the dynamics of relativistic nuclear collisions. Such collisions are 
presently under intense experimental investigation at CERN's SPS and 
Brookhaven National Laboratory's AGS and (beginning Fall 1999) 
RHIC accelerators. 

In this set of
lectures I give an overview over the basic concepts and notions
of relativistic fluid dynamics as applied to the physics of heavy-ion
collisions. The aim is not to present a detailed and complete
review of this field, but to provide a foundation to understand
the literature on current research activities in this field.
This has the consequence that the list of references
is far from complete, that I will not make any attempt to compare to actual
experimental data, and that some interesting, but more applied 
topics (such as transverse collective flow) will not be discussed here.
In Section 2, I discuss the basic concepts of relativistic
fluid dynamics. First, I present a derivation of the fluid-dynamical
equations of motion. A priori, there are more unknown functions 
than there are equations, and one has to devise approximation
schemes in order to close the set of equations of motion.
The most simple is the ideal fluid approximation, which simply discards
some of the unknown functions. Another one is the assumption of
small deviations from local thermodynamical equilibrium, which leads
to the equations of dissipative fluid dynamics. A brief discussion of
multi-fluid models concludes this section. In Section 3 I discuss 
numerical aspects of solution schemes for ideal relativistic
fluid dynamics. Section 4 is devoted to a discussion of one-dimensional 
solutions of ideal fluid dynamics. After presenting a classification of
possible wave patterns in one spatial dimension, for both thermodynamicall
normal as well as anomalous matter, I discuss
the expansion of semi-infinite matter into vacuum. This naturally leads
to the discussion of the Landau model for the one-dimensional
expansion of a finite slab of matter. The Landau model is historically
the first fluid-dynamical model for relativistic nuclear collisions. 
However, more realistic is, at least for ultrarelativistic collision 
energies, the so-called the Bjorken model which is subsequently
presented. The main result of this section is the time delay in the expansion
of the system due to the softening of the equation of state in
a phase transition region. This may have potential experimental consequences
for nuclear collisions at RHIC energies, where one wants to study the
transition from hadron to quark and gluon degrees of freedom.
Finally, Section 5 concludes this set of lectures 
with a discussion on how to decouple particles
from the fluid evolution in the so-called ``freeze-out'' process and
compute experimentally observable quantities like single inclusive
particle spectra.

Units are $\hbar = c = k_{\rm B} = 1$.
The metric tensor is $g^{\mu \nu} = \mbox{diag} (+,-,-,-)$.
Upper greek indices are contravariant, lower greek indices covariant.
The scalar product of two 4-vectors $a^\mu\, , \,\, b^\mu$ is denoted by 
$a^\mu\, g_{\mu \nu}\, b^\nu \equiv a^\mu \, b_\mu \equiv a \cdot b$.
The latter notation is also used 
for the scalar product of two 3-vectors $\vec{a},\, \vec{b}$, 
$\vec{a} \cdot \vec{b}$.

\section{The Basics} 

In this section, I first derive the conservation equations of relativistic
fluid dynamics. If there are $n$ conserved charges in the fluid, there
are $4+n$ conservation equations: 1 for the conservation of energy,
3 for the conservation of 3-momentum, and $n$ for the conservation
of the respective charges. 
In the general case, however, there are $10+4\,n$ independent
variables: the 10 independent components of the energy-momentum tensor
(which is a symmetric tensor of rank 2),
and the 4 independent components of the 4-vectors of the $n$ charge currents.
Thus, the system of fluid-dynamical equations is not closed, and one
requires an approximation in order to solve it. 

The simplest approximation is the ideal fluid assumption which reduces 
the number of unknown functions to $5+n$. The equation of state of the
fluid then provides the final equation to close the system of conservation
equations and to solve it uniquely. 
Another approximation is the assumption of small deviations from
an ideal fluid and leads to the equations of dissipative fluid dynamics.
In this approximation one provides an additional set of $6 + 3\, n$
equations to close the set of equations of motion.
Finally, I also briefly discuss multi-fluid-dynamical models.

\subsection{The Conservation Equations}

Fluid dynamics is equivalent to the conservation of energy, momentum, and net
charge number. Consider a single fluid characterized locally in space-time
by its energy-momentum tensor $T^{\mu \nu} (x)$ and by the $n$ conserved
net charge currents $N^\mu_i(x)\, ,\, i=1,\ldots,n$. (Conserved charges
are for example the electric charge, baryon number, strangeness, charm,
etc.)
Consider now an arbitrary hypersurface $\Sigma$ in 4-dimensional space-time.
The tangent 4-vector on this surface is $\Sigma^\mu(x)$. 
The normal vector on a surface element $\D \Sigma$ of 
$\Sigma$ is denoted by $\D \Sigma_\mu(x)$. By definition, $\D \Sigma \cdot
\Sigma = 0$. The amount of net charge of type $i$ and of energy and momentum
flowing through the surface element $\D \Sigma$ is given by 
\begin{eqnarray}
\D N_i & \equiv & \D \Sigma \cdot N_i\,\, ,\,\,\, i=1,\ldots, n \enspace , \\
\D P^\mu & \equiv & \D \Sigma_\nu \, T^{\mu \nu} \,\, ,\,\,\, \mu = 0,\ldots,3
\enspace .
\end{eqnarray}
Now consider an arbitrary space-time volume $V_4$ with a
{\em closed\/} surface $\Sigma$. If there are neither sources nor sinks
of net charge and energy-momentum inside $\Sigma$, one has
\begin{eqnarray}
\oint_\Sigma \D \Sigma \cdot N_i & \equiv & 0 
\,\, ,\,\,\, i=1,\ldots, n\enspace, \\
\oint_\Sigma \D \Sigma_\nu \, T^{\mu \nu} & \equiv & 0 
\,\, ,\,\,\, \mu = 0,\ldots,3 \enspace.
\end{eqnarray}
Gauss theorem then leads immediately to the {\em global conservation
of net charge and energy-momentum}:
\begin{eqnarray} \label{ris:globalcons1}
\int_{V_4} \D^4 x\,  \partial_\mu N_i^\mu & \equiv & 0
\,\, ,\,\,\, i=1,\ldots, n  \enspace, \\
\int_{V_4} \D^4 x\, \partial_\nu T^{\mu \nu} & \equiv & 0 
\,\, ,\,\,\, \mu = 0,\ldots,3 \enspace.
\label{ris:globalcons2}
\end{eqnarray}
For arbitrary $V_4$, however, one has to require that the integrands in
(\ref{ris:globalcons1},\ref{ris:globalcons2}) vanish, which leads to
{\em local conservation of net charge and energy-momentum}:
\begin{eqnarray} \label{ris:localcons1}
\partial_\mu N_i^\mu & \equiv & 0 \,\, ,\,\,\, i=1,\ldots, n  \enspace, \\
\partial_\mu T^{\mu \nu} & \equiv & 0 
\,\, ,\,\,\, \nu = 0,\ldots,3 \enspace.
\label{ris:localcons2}
\end{eqnarray}
These are the equations of motion of relativistic fluid dynamics
\cite{ris:LL}. Note that there
are $4+n$ equations, but $10 + 4\, n$ independent unknown functions
$T^{\mu \nu}(x)\, ,\,\, N_i^{\mu}(x)$. ($T^{\mu\nu}$ is a symmetric
tensor of rank 2 and therefore has 10 independent components, the
$N_i^{\mu}$ are 4-vectors with 4 independent components.) Therefore,
the system of fluid-dynamical equations is a priori not closed and
cannot be solved in complete generality. 
One requires additional assumptions to
close the set of equations. One possibility is to reduce the number of
unknown functions, another is to provide $6+3\, n$
additional equations of motion which determine all unknown functions
uniquely. Both possibilities will be discussed in the following 
subsections.

\subsection{Tensor Decomposition and Choice of Frame}

Before discussing approximations to close the system of conservation
equations, it is convenient to perform a
tensor decomposition of $N_i^\mu \, , T^{\mu\nu}$ with respect to
an {\em arbitrary, time-like, normalized\/} 4-vector $u^\mu$, 
$u \cdot u = 1$. The projector onto the 3-space orthogonal to
$u^\mu$ is denoted by
\begin{equation}
\Delta^{\mu \nu} \equiv g^{\mu \nu} - u^\mu u^\nu\, ,\,\,
\Delta^{\mu \nu} u_\nu = 0 \, , \,\, \Delta^{\mu \alpha} 
\Delta_{\alpha}^\nu = \Delta^{\mu \nu} \enspace.
\end{equation}
Then the tensor decomposition reads:
\begin{eqnarray}
N^\mu_i & = & n_i \, u^\mu + \nu_i^\mu \enspace , \\
T^{\mu \nu} & = & \epsilon\, u^\mu u^\nu - p\, \Delta^{\mu \nu}
+ q^\mu u^\nu + q^\nu u^\mu + \pi^{\mu \nu} \enspace,
\end{eqnarray}
where
\begin{equation}
n_i \equiv N_i \cdot u
\end{equation}
is the {\em net density\/} of charge of type $i$ in the frame where
$u^\mu = (1,\vec{0})$ (subsequently denoted as the {\em local
rest frame}, LRF),
\begin{equation}
\nu_i^\mu \equiv \Delta^{\mu}_\nu N_i^\nu 
\end{equation}
is the {\em net flow\/} of charge of type $i$ in the LRF,
\begin{equation}
\epsilon \equiv u_\mu T^{\mu \nu} u_\nu
\end{equation}
is the {\em energy density\/} in the LRF,
\begin{equation} \label{ris:isotropicpress}
p \equiv - \frac{1}{3} T^{\mu \nu} \Delta_{\mu \nu}
\end{equation}
is the {\em isotropic pressure\/} in the LRF,
\begin{equation}
q^{\mu} \equiv \Delta^{\mu \alpha} T_{\alpha \beta} u^{\beta}
\end{equation}
is the {\em flow of energy\/} or {\em heat flow\/} in the LRF, and
\begin{equation}
\pi^{\mu \nu} \equiv \left[ \frac{1}{2} \left( \Delta^\mu_\alpha
\Delta_\beta^\nu + \Delta^\mu_\beta \Delta_\alpha^\nu \right)
- \frac{1}{3} \Delta^{\mu \nu} \Delta_{\alpha \beta} \right]\,
T^{\alpha \beta} \label{ris:stresstens}
\end{equation}
is the {\em stress tensor\/} in the LRF. Note that the particular
projection (\ref{ris:stresstens}) is {\em trace-free}. (The trace of
the projection $\Delta^\mu_\alpha T^{\alpha \beta} \Delta_\beta^\nu$ is
absorbed in the definition of $p$.)
The tensor decomposition replaces the original $10 + 4\, n$ unknown
functions by an equal number of new unknown functions
$n_i \, (n),\, \nu_i^{\mu} \, (3\, n), \, \epsilon \, (1)$, $ p \, (1)$,
$q^{\mu}\, (3),$ and $\pi^{\mu \nu}$ (5). 

So far, $u^\mu$ is arbitrary. However, one can give it a physical
meaning by choosing it either to be
\begin{equation}
u_{\rm E}^\mu \equiv \frac{N_i^\mu}{\sqrt{N_i \cdot N_i}} \enspace ,
\end{equation}
or (which is an implicit definition)
\begin{equation}
u_{\rm L}^\mu \equiv \frac{T^\mu_\nu u_{\rm L}^\nu}{\sqrt{u_{\rm L}^\alpha 
T_\alpha^\beta T_{\beta \gamma} u_{\rm L}^\beta}} \enspace.
\end{equation}
The first choice means that $u_{\rm E}^\mu$ is the physical 
4-velocity of the {\em flow\/} 
of net charge $i$. The LRF is then the {\em local rest frame of
the flow of net charge\/} $i$, 
i.e., the frame where $N_i^\mu = (N_i^0, \vec{0})$.
In this frame, there is obviously no flow of charge $i$, 
$\nu_i^\mu \equiv 0$, and $N_i^0 \equiv n_i$. This LRF 
is called {\em Eckart frame}. Note, however, that not all net charges
need to flow with the same velocity, $\nu_j^\mu$ might be $\neq 0$
for $j \neq i$. The number of unknown functions is still
$10 + 4\, n$, since the 3 previously unknown functions $\nu_i^\mu$ have 
been merely replaced by the 3 independent components of $u_{\rm E}^\mu$ 
($u_{\rm E} \cdot u_{\rm E} = 1$!), which now have to be determined 
dynamically from the conservation equation for $N_i^\mu$.

The second choice means that $u_{\rm L}^\mu$ is the physical 4-velocity
of the {\em energy flow}. The LRF is the {\em local rest frame
of the energy flow}. It is obvious that in this frame $q^\mu \equiv 0$.
This frame is called {\em Landau frame}. The number of 
unknown functions is still $10 + 4\, n$, since the 3 previously unknown 
functions $q^\mu$ have been merely replaced by the 3 independent 
components of $u_{\rm L}^\mu$ ($u_{\rm L} \cdot u_{\rm L} = 1$!), 
which now have to be determined dynamically from 
the conservation equation for $T^{\mu \nu}$.
Other choices of rest frames are also possible, for a discussion, see
\cite{ris:deGroot}.

\subsection{Ideal Fluid Dynamics}

Consider an ideal gas in {\em local thermodynamical
equilibrium}. The sin\-gle-par\-ti\-cle phase space distribution 
for fermions or bosons then reads
\begin{equation} \label{ris:feq}
f_0(k,x) = \frac{g}{(2 \pi)^3} \, 
\frac{1}{\exp{(k\cdot u(x) - \mu(x))/T(x)} \pm 1}\enspace ,
\end{equation}
where $u^{\mu}(x)$ is the {\em local\/} average 4-velocity of the
particles, $\mu(x)$ and $T(x)$ are {\em local\/} chemical potential and
temperature, and $g$ counts internal degrees of freedom (spin, isospin,
color, etc.) of the particles.
The chemical potential of the particles is defined
as $\mu \equiv \sum_{i=1}^n q_i \mu_i$, where $\mu_i$ are the chemical
potentials which control the net number of charge of type $i$, and
$q_i$ is the individual charge of type $i$ carried by a particle.
The chemical potential for antiparticles is $\bar{\mu} = - \mu$
(in thermodynamical equilibrium). Let us define the single-particle
phase space distribution for antiparticles by $\bar{f}_0(\bar{\mu}) 
= f_0(-\mu)$.

The kinetic definitions of the net current of charge of type $i$ 
and of the energy-momentum tensor are \cite{ris:deGroot}
\begin{eqnarray} \label{ris:Nmukin}
N_i^\mu(x) &\equiv& q_i \int \frac{\D^3 \vec{k}}{E} \, k^\mu\, 
\left[f_0(k,x) - \bar{f}_0(k,x) \right] \enspace, \\
T^{\mu \nu}(x) &\equiv& \int \frac{\D^3 \vec{k}}{E}\,
k^\mu k^\nu \, \left[ f_0(k,x) + \bar{f}_0(k,x) \right]\enspace,
\label{ris:Tmunukin}
\end{eqnarray}
where $E \equiv \sqrt{\vec{k}^2 + m^2}$ is the on-shell energy of the
particles and $m$ their rest mass. Inserting (\ref{ris:feq}) one computes
\begin{eqnarray} \label{ris:Ni}
N_i^\mu & = & n_i \, u^\mu \enspace , \\
T^{\mu \nu} & = & \epsilon\, u^\mu u^\nu - p \, \Delta^{\mu \nu} \enspace,
\label{ris:Tmunu}
\end{eqnarray}
where
\begin{equation}
n_i \equiv g\,q_i \int \frac{\D^3 \vec{k}}{(2 \pi)^3}\, 
\left[n(E) - \bar{n}(E)\right]
\end{equation}
is the {\em thermodynamic\/} net number density of charge of type $i$
of an {\em ideal gas},
and the Fermi--Dirac or Bose--Einstein
distribution was denoted by $n(E) \equiv 1/(\exp[(E-\mu)/T] \pm 1)$, 
$\bar{n}(E) \equiv 1/(\exp[(E+\mu)/T] \pm 1)$. Furthermore,
\begin{equation}
\epsilon \equiv g \int \frac{\D^3 \vec{k}}{(2 \pi)^3}\, 
E\, \left[n(E) + \bar{n}(E)\right]
\end{equation}
is the {\em thermodynamic ideal gas\/} energy density, and
\begin{equation} \label{ris:idgaspress}
p \equiv g \int \frac{\D^3 \vec{k}}{(2 \pi)^3}\, 
\frac{\vec{k}^2}{3\, E} \left[n(E) + \bar{n}(E)\right]
\end{equation}
is the {\em thermodynamic ideal gas\/} pressure.
The form (\ref{ris:Ni},\ref{ris:Tmunu}) implies that for an ideal gas in
local thermodynamical equilibrium the functions
$\nu_i^\mu = q^\mu = \pi^{\mu \nu} = 0$, i.e., there is no
flow of charge or heat with respect to the particle flow velocity $u^\mu$,
and there are no stress forces. This implies furthermore (and can be
confirmed by an explicit calculation) that for an ideal gas in
local thermodynamical equilibrium $u_{\rm E}^\mu \equiv 
u_{\rm L}^\mu \equiv u^\mu$,
i.e., Eckart's and Landau's choice of frame coincide with the local rest
frame of particle flow.

This consideration of an ideal gas in local thermodynamical equilibrium
serves as a motivation for the so-called {\em ideal fluid approximation}. 
In this approximation, one starts on the {\em macroscopic\/} level of
fluid variables $N_i^\mu, \, T^{\mu \nu}$ and {\em a priori\/} takes them
to be of the form (\ref{ris:Ni}) and (\ref{ris:Tmunu}). 
The corresponding fluid is referred to as an
{\em ideal fluid}. Without any further assumption, however, the corresponding
system of $4+n$ equations of motion contains $5+n$ unknown functions,
$\epsilon, p, u^\mu$, and $n_i,\, i=1,\ldots,n$. One therefore has to
specify an {\em equation of state\/} for the fluid, for instance (and most
commonly taken) of the form $p(\epsilon,n_1,\ldots,n_n)$. This
closes the system of equations of motion.
 
The equation of state is the {\em only\/} place where 
information enters about the nature of the particles in the fluid and the 
{\em microscopic\/}
interactions between them. Usually, the equation of state for the
fluid is taken to be the {\em thermodynamic\/} equation of state, as 
computed for a system in {\em thermodynamical equilibrium}.
The process of closing the system of equations of motions 
by assuming a thermodynamic equation of state
therefore involves the {\em implicit assumption\/}
that the {\em fluid is in local thermodynamical equilibrium}.
It is important to note, however, that 
the {\em explicit form\/} 
of the equation of state is {\em completely unrestricted\/}, 
for instance it can have anomalies like phase transitions. 

The ideal fluid approximation therefore allows to consider a wider class
of systems than just an ideal gas
in local thermodynamical equilibrium, which served as a motivation for this
approximation. An ideal gas has a very specific equation of state without
any anomalies and is given by (\ref{ris:idgaspress})
which defines $p(T,\mu_1,\ldots,\mu_n)$ (which in turn allows to
determine all other thermodynamic functions from the first law 
and the fundamental relation of thermodynamics, and thus to specify
$p(\epsilon,n_1,\ldots,n_n)$, see the following remarks). 

I close this subsection with three remarks. The first concerns the notion
of an equation of state which is {\em complete\/} 
in the thermodynamic sense.
Such an equation of state allows (by definition) to determine, for
given values of the independent thermodynamic variables, all
other thermodynamic functions from the first law of thermodynamics
(or one of its Legendre transforms) 
\begin{equation} \label{ris:1stlaw}
\D s = \frac{1}{T}\, \D \epsilon - \sum_{i=1}^n \frac{\mu_i}{T} \, n_i
\enspace ,
\end{equation}
$s$ being the entropy density, and from
the fundamental relation of thermodynamics
\begin{equation} \label{ris:fundamental}
\epsilon + p = T\, s + \sum_{i=1}^n \mu_i\, n_i \enspace.
\end{equation}
Obviously, for independent thermodynamic variables 
$\epsilon,\, n_1, \ldots, \, n_n$, an equation of state of the form
$s(\epsilon,n_1,\ldots,n_n)$ is complete in this sense, since partial
differentiation of this function yields, from (\ref{ris:1stlaw}),
the functions $1/T$, $\mu_1/T$, $\ldots$, $\mu_n/T$. Then, the
fundamental relation (\ref{ris:fundamental}) yields the last unknown
thermodynamic function, $p$.

Another example of a complete equation of state is $p(T,\mu_1,\ldots,\mu_n)$,
since the (multiple) Legendre transform of (\ref{ris:1stlaw}) reads
\begin{equation}
\D p = s\,  \D T + \sum_{i=1}^n n_i \, \D \mu_i
\end{equation}
(which is also known as the Gibbs--Duhem relation), such that the 
thermodynamic functions $s,\, n_1, \ldots, \, n_n$ can be determined
from partial differentiation of $p$. The last unknown thermodynamic
function, $\epsilon$, can then be determined from (\ref{ris:fundamental}).

The equation of state $p(\epsilon, n_1,\ldots,n_n)$ is, however, 
{\em not a complete\/} equation of state in the thermodynamic sense.
Partial differentiation of this function yields thermodynamic functions
$\partial p/\partial \epsilon, \, \partial p/\partial n_i, \, i =1,\ldots, n$,
which in general do {\em not allow\/} to infer the values of $T,\, s,$
and $\mu_i$, $i=1, \ldots,n$.

The second remark concerns the assumption of local thermodynamical 
equilibrium. In order to achieve local thermodynamical equilibrium,
spatio-temporal variations of the macroscopic fluid fields have to be small
compared to microscopic reaction rates which drive the system (locally)
towards thermodynamical equilibrium. A quantity that characterizes 
spatio-temporal variations of the macroscopic fields is the so-called
{\em expansion scalar\/} $\theta \equiv \partial \cdot u$. It determines
the (local) rate of expansion of the fluid. Microscopic reaction rates 
are essentially given by the product of cross section and local particle
density, $\Gamma \simeq \sigma\, n$. The criterion for local thermodynamical
equilibrium then reads
\begin{equation}
\Gamma \gg \theta \, , \,\,\, \mbox{or} \,\,\,\, \sigma \gg \theta/n \enspace .
\end{equation}

The third remark concerns entropy production. In ideal fluid dynamics,
the entropy current is defined as
\begin{equation} \label{ris:ent}
S^\mu  \equiv s \, u^\mu \enspace.
\end{equation}
Taking the projection of energy-momentum conservation in the direction
of $u_\nu$ one derives
\begin{equation}
0 = u_\nu \, \partial_\mu T^{\mu \nu} = \dot{\epsilon} + (\epsilon + p)\, 
\theta \enspace ,
\end{equation}
where $\dot{a} \equiv u \cdot \partial\, a$ is a comoving time derivative and
where use has been made of the fact that $u^\mu$ is normalized, i.e.,
$\partial_\mu (u \cdot u) = 0$. With the first law of thermodynamics
(\ref{ris:1stlaw}) and the fundamental relation of thermodynamics 
(\ref{ris:fundamental}) one rewrites this as
\begin{equation}
T \, (\dot{s} +s\, \theta) + \sum_{i=1}^n \mu_i \, (\dot{n}_i + n_i \,
\theta) = 0 \enspace .
\end{equation}
Finally, employing net charge conservation 
$\partial \cdot N_i \equiv \dot{n}_i + n_i\, \theta = 0$ yields
\begin{equation} \label{ris:entropycons}
\dot{s} + s\, \theta \equiv \partial \cdot S =0 \enspace,
\end{equation}
i.e., the entropy current is conserved in ideal fluid dynamics.
As we shall see in one of the following section, however, this proof only
holds where the partial derivatives in these equations are 
well-defined, i.e., for continuous solutions of ideal fluid dynamics. 
Discontinuous solutions will in fact be shown to produce entropy.

\subsection{Dissipative Fluid Dynamics}

In dissipative fluid dynamics one does not set 
$\nu_i^\mu, \, q^\mu, \, \pi^{\mu \nu}$ a priori to zero, but specifies
them through additional equations. There are two ways to obtain
the latter. The first is phenomenological and starts from the second law
of thermodynamics, i.e., the principle of non-decreasing entropy,
\begin{equation}
\partial \cdot S \geq 0 \enspace .
\end{equation}
The second way resorts to kinetic theory to derive the respective equations.
In principle, both ways require the additional assumption that 
deviations from local thermodynamical equilibrium are small.
To make this statement more concise, let us introduce the {\em equilibrium
pressure\/} $p_{\rm eq} = p_{\rm eq}(\epsilon, n_1, \ldots, n_n)$, i.e., it is
the pressure as computed from the equation of state for given values of
$\epsilon, \,n_1, \ldots, \, n_n$. In a general non-equilibrium
(dissipative) situation, however, $p_{\rm eq}$ is different from
the isotropic pressure $p$ defined through (\ref{ris:isotropicpress}). 
Denote the difference by $\Pi \equiv p_{\rm eq}-p$.
Then, the requirement that deviations from local thermodynamical equilibrium
are small is equivalent to requiring $\nu_i^\mu,\, q^\mu, \pi^{\mu \nu}$,
and $\Pi$ to be small compared to $\epsilon, \, p_{\rm eq}$, and $n_i$.

I first outline the phenomenological approach to derive
the equations of dissipative fluid dynamics. For the sake of definiteness, 
in the remainder of this subsection let us 
consider a system of one particle species only and let us assume that
the total {\em particle number\/} of this species is conserved 
(implying that no annihilation or creation processes take place, i.e.,
we do not consider the corresponding antiparticles). The particle
number current then replaces the net charge current.
We shall also work in the Eckart frame, where $\nu^\mu \equiv 0$.
Let us make an Ansatz for the entropy 4-current $S^\mu$. 
In the limit of vanishing $q^\mu, \, \pi^{\mu \nu}$, and $\Pi$, 
the entropy 4-current
should reduce to the one of ideal fluid dynamics, $S^\mu \rightarrow s \,
u^\mu$. The only non-vanishing 4-vector which can be formed from
the available tensors $u^\mu,\, q^\mu,$ and $\pi^{\mu \nu}$ is
$\beta \, q^\mu$, where $\beta$ is an arbitrary coefficient (remember
$\pi^{\mu \nu} u_\nu =0$). Therefore,
\begin{equation} \label{ris:Ansatz1}
S^\mu = s\, u^\mu + \beta\, q^\mu \enspace.
\end{equation}
With this Ansatz one computes with the help of $u_\nu \partial_\mu T^{\mu \nu}
= 0$ and $\partial \cdot N = \dot{n} + n \, \theta=0$:
\begin{equation}
T \partial \cdot S = (T\beta -1) \partial \cdot q + q \cdot (\dot{u} + T
\partial \beta) + \pi^{\mu \nu} \partial_\mu u_\nu + \Pi \, \theta \geq 0
\enspace .
\end{equation}
The simplest way to ensure this inequality is to choose
\begin{eqnarray}
\beta & \equiv & 1/T \enspace, \\
\Pi & \equiv & \zeta \, \theta \enspace, \\
q^\mu & \equiv &  \kappa \, T\, \Delta^{\mu \nu} \,(\partial_\nu \ln T
- \dot{u}_\nu) \enspace , \\
\pi^{\mu \nu} & \equiv & 2\, \eta \left[ \frac{1}{2} \left( \Delta^\mu_\alpha
\Delta_\beta^\nu + \Delta^\mu_\beta \Delta_\alpha^\nu \right)
- \frac{1}{3} \Delta^{\mu \nu} \Delta_{\alpha \beta} \right]\, \partial^\alpha
u^\beta \enspace,
\end{eqnarray}
where $\zeta, \, \eta$, and $\kappa$ are the (positive) {\em bulk
viscosity, shear viscosity\/} and {\em thermal conductivity\/}
coefficients. Note that these equations define the dissipative
corrections as {\em algebraic\/} functions of gradients of 
the flow velocity $u^\mu$ and the equilibrium temperature $T$.
With these choices, 
\begin{equation}
\partial \cdot S = \frac{\Pi^2}{\zeta \,T} - \frac{q \cdot q}{\kappa\, T^2}
+ \frac{\pi^{\mu \nu}  \pi_{\mu \nu}}{2\, \eta \,T}\enspace ,
\end{equation}
which is obviously larger or equal to zero (remember that $q \cdot q <0$,
which can be most easily proven from $q \cdot u =0$ 
in the frame where $u^\mu = (1,\vec{0})$). While this ensures
the second law of thermodynamics, it was shown \cite{ris:hiscock} that the
resulting equations of motion are {\em unstable\/} under perturbations
and support {\em acausal\/}, i.e., {\em superluminous\/} propagation of
information. They are therefore not suitable as candidates for a
{\em relativistic\/} theory of dissipative fluid dynamics.

A solution to this dilemma was presented by M\"uller \cite{ris:muller}, 
and Israel and Stewart \cite{ris:israelstewart}. They observed that the Ansatz
(\ref{ris:Ansatz1}) for the entropy current should not only contain 
first order terms in the dissipative corrections, but also second order 
terms:
\begin{equation}
S^\mu = s \, u^\mu + \beta \, q^\mu + Q^\mu \enspace,
\end{equation}
where
\begin{equation}
Q^\mu \equiv \alpha_0\, \Pi \, q^\mu + \alpha_1\, \pi^{\mu \nu}\, q_\nu
+ u^\mu \, \left( \beta_0\, \Pi^2 + \beta_1\, q \cdot q + \beta_2\,
\pi^{\nu \lambda} \pi_{\nu \lambda} \right) 
\end{equation}
is second order in the dissipative quantities $\Pi, \, q^\mu,$ and 
$\pi^{\mu \nu}$. Inserting this into $\partial \cdot S \geq 0$ leads
to {\em differential equations\/} for $\Pi, \, q^\mu$, and $\pi^{\mu \nu}$
which involve the coefficients $\zeta,\, \eta, \, \kappa,\, \alpha_0,\,
\alpha_1,\, \beta_0,\, \beta_1,\, \beta_2$. It can be shown that, for
reasonable values of these coefficents, the resulting 14 equations of motion
(the 9 equations that determine $\Pi,\, q^\mu$, and $\pi^{\mu \nu}$ and
the 5 conservation equations for $N^\mu,\, T^{\mu \nu}$) 
are stable and causal. 

In the phenomenological approach, the values of these coefficients
are not determined. In the second approach, however, based on
kinetic theory, they can be explicitly computed along with deriving the
additional 9 equations of motion for $\Pi,\, q^\mu,$ and $\pi^{\mu \nu}$.
This will be outlined in the following.

Let us start by writing the single-particle phase space distribution in
local equilibrium (\ref{ris:feq}) as
\begin{equation}
f_0 (k,x) = \frac{g}{(2 \pi)^3} \, \left[ \exp\{y_0(k,x)\} \pm 1 \right]^{-1}
\enspace ,
\end{equation}
where $y_0 (k,x) \equiv [k \cdot u(x) - \mu(x)]/T(x)$.
Now assume that the {\em non-equilibrium\/} phase space distribution,
written in the form
\begin{equation}
f(k,x) \equiv \frac{g}{(2 \pi)^3} \, \left[ \exp\{y(k,x)\} \pm 1 \right]^{-1}
\enspace ,
\end{equation}
deviates only slightly from the equilibrium distribution function
$f_0(k,x)$, or in other words:
\begin{equation}
y(k,x) \simeq y_0(k,x) + \varepsilon_1(x) + k \cdot \varepsilon_2(x) +
k_\mu k_\nu \, \varepsilon_3^{\mu \nu}(x) \enspace,
\end{equation}
where $\varepsilon_1(x),\, \varepsilon_2^\mu (x)$, and $\varepsilon_3^{\mu
\nu}$ are small compared to $T(x),\, \mu(x)$. Then one can expand
$f(k,x)$ around $f_0(k,x)$ to first order in these small quantities:
\begin{equation}
f(k,x) \simeq f_0(k,x) \left( 1 + \left[ 1 \mp \frac{(2 \pi)^3}{g}\,
f_0(k,x) \right] \, \left[ y(k,x)- y_0(k,x) \right] \right) \enspace . 
\end{equation}
Note that $f(k,x)$ depends on the 14 variables
$\mu/T - \varepsilon_1,\, u^\mu/T + \varepsilon_2^\mu,$ and
$\varepsilon_3^{\mu\nu}$. ($\varepsilon_3^{\mu\nu}$ is a symmetric
tensor of rank 2, and therefore naively has 10 independent components.
However, its trace can be absorbed in a redefinition of the first
variable $\mu/T- \varepsilon_1$, therefore it actually has only 9 independent
components.) 

Inserting $f(k,x)$ into the kinetic theory
definition of $N^\mu$ and $T^{\mu \nu}$, (\ref{ris:Nmukin}) 
and (\ref{ris:Tmunukin}),
(with $f_0$ replaced by $f$ and, since we do not consider
antiparticles, discarding $\bar{f}_0$), one can establish relations between 
the 14 unknown {\em macroscopic\/}  functions (in the Eckart frame) 
$\epsilon,\, n,\, u^\mu,\, \Pi,\, q^{\mu},\, \pi^{\mu\nu}$ and the
14 variables $\mu/T-\varepsilon_1,\, u^\mu/T+\varepsilon_2^\mu,\,
\varepsilon_3^{\mu\nu}$. This uniquely determines the 
non-equilibrium single-particle phase space distribution $f(k,x)$ in terms
of the macroscopic, i.e., fluid-dynamical variables. This identification 
involves one subtlety: as in ideal fluid dynamics one still has to know
the value of the (equilibrium) pressure $p_{\rm eq}$ to determine all unknown
quantities. The equilibrium pressure $p_{\rm eq}$ is, however, only
known as a function of the {\em equilibrium\/} energy density 
$\epsilon_0$ and the {\em equilibrium\/} particle number density $n_0$, 
but not as function of the actual energy density $\epsilon$ and
particle number density $n$. Two additional assumptions are required, 
namely that
\begin{eqnarray} \label{ris:ass1}
\epsilon \equiv u_\mu T^{\mu\nu} u_\nu & = & 
\epsilon_0 \equiv u_\mu T_0^{\mu\nu} u_\nu \enspace , \\
n \equiv u \cdot N & = & n_0 \equiv u \cdot N_0 \enspace , \label{ris:ass2}
\end{eqnarray}
where $T_0^{\mu\nu}$ and $N_0^\mu$ are the (kinetic) energy-momentum tensor and
particle number current computed with the {\em equilibrium\/} phase space 
distribution
$f_0(k,x)$. Then $p_{\rm eq}(\epsilon,n) \equiv p_{\rm eq}(\epsilon_0,n_0)$
and the value of the equilibrium pressure $p_{\rm eq}$ is also determined.
On close inspection, these additional assumptions do not pose
any further restriction on the set of 14 unknown functions, but merely serve
as definitions of (equilibrium) temperature $T$ and chemical potential
$\mu$ corresponding to a given energy density $\epsilon$ and particle number
density $n$. Another way to say this is that the assumptions (\ref{ris:ass1}), 
(\ref{ris:ass2})
determine a  local equilibrium phase space distribution $f_0(k,x)$. 
However, in a non-equilibrium context this distribution has no actual
dynamical meaning, and one is therefore free to choose it in a way which
fulfills (\ref{ris:ass1}) and (\ref{ris:ass2}).

The next step consists of deriving the equations of motion for the
14 unknown functions $\epsilon,\, n,\, u^\mu, \, \Pi,\, q^\mu,\, \pi^{\mu\nu}$.
To this end, one takes the first {\em three\/} moments of the Boltzmann
equation for $f(k,x)$,
\begin{equation}
k \cdot \partial \, f(k,x) = {\cal C} [f] \enspace .
\end{equation}
This results in
\begin{eqnarray} 
\int \frac{\D^3 \vec{k}}{E}\, k \cdot \partial \, f(k,x) 
\equiv  \partial \cdot N 
& = &  \int \frac{ \D^3 \vec{k}}{E}\, {\cal C}[f]
\equiv 0 \enspace , \label{ris:Nmu2} \\
\int \frac{\D^3 \vec{k}}{E}\, k^\mu k^\nu\, \partial_\mu \, f(k,x) 
\equiv  \partial_\mu T^{\mu \nu}
& = & \int \frac{ \D^3 \vec{k}}{E}\, k^\nu \,
{\cal C}[f] \equiv 0 \enspace , \label{ris:Tmunu2} \\
\int \frac{\D^3 \vec{k}}{E}\, k^\mu k^\nu k^\lambda \, \partial_\mu 
\, f(k,x) \equiv \partial_\mu S^{\mu \nu \lambda}
& = &  \int \frac{ \D^3 \vec{k}}{E}\, k^\nu k^\lambda \,
 {\cal C}[f] \equiv X^{\nu \lambda} \enspace . \label{ris:Smunulam} 
\end{eqnarray}
Note that conservation of particle number, energy, and momentum leads to 
vanishing right-hand sides for eqs.\ (\ref{ris:Nmu2}) and (\ref{ris:Tmunu2}). 
The structure of the microscopic collision term ${\cal C}$ is such 
that these requirements are fulfilled (particle number and energy-momentum
conservation in microscopic collisions between particles) \cite{ris:deGroot}. 
On the other hand, the right-hand
side of Eq.\ (\ref{ris:Smunulam}) does not vanish, 
since there is no corresponding
microscopic conservation law. Note that the trace of (\ref{ris:Smunulam})
is equivalent to $m^2$ times Eq.\ (\ref{ris:Nmu2}), such that $X^\nu_\nu 
\equiv 0$. Therefore,
only 9 out of the set of 10 equations (\ref{ris:Smunulam}) are independent.
Together with the 5 equations (\ref{ris:Nmu2}) and (\ref{ris:Tmunu2}), 
these 9 equations determine the set of 14 unknown functions of dissipative
fluid dynamics. The 9 independent equations (\ref{ris:Smunulam}) are equivalent
to the 9 equations derived from $\partial \cdot S \geq 0$ in the 
phenomenological approach. The 
unknown phenomenological coefficents $\zeta,\, \kappa,\, \eta,\,
\alpha_0,\, \alpha_1,\, \beta_0,\, \beta_1,$ and $\beta_2$ can now
be explicitly identified from suitable projections of $X^{\nu\lambda}$.
Israel and Stewart have shown 
\cite{ris:israelstewart} that the resulting equations fulfill the
requirements of hyperbolicity and causality.

This concludes the brief survey of dissipative fluid dynamics. So far,
no serious attempt has been made to apply relativistic
dissipative fluid dynamics towards the description of heavy-ion collisions.
First steps were done by
Mornas and Ornik \cite{ris:mornas} who investigated the broadening of 
collisional
shock waves through dissipative effects in a simple one-dimensional geometry.
Also, Prakash et al.\ generalized the Israel--Stewart theory to a mixture
of several particle species \cite{ris:prakash}.

\subsection{Multi-fluid Dynamics}

In multi-fluid dynamics one considers not a single, but several
fluids $j=1,\ldots,M$, characterized by the net charge currents
$N_{ij}^\mu$ (the net current of conserved charge $i$ in fluid $j$) and
energy-momentum tensors $T_j^{\mu\nu}$. There is overall net charge
and energy-momentum conservation,
\begin{eqnarray} \label{ris:Nmuj}
\partial \cdot N_i & = & 0\,\,\, , \,\,\,\, N_i^\mu \equiv \sum_{j=1}^M
N_{ij}^\mu \enspace , \\
\partial_\mu T^{\mu \nu} & = & 0 \,\,\, , \,\,\,\, T^{\mu \nu} 
\equiv \sum_{j=1}^M T_j^{\mu \nu} \enspace ,  \label{ris:Tmunuj}
\end{eqnarray}
but not for each fluid separately,
\begin{equation}
\partial \cdot N_{ij} = S_{ij} \,\,\, , \,\,\,\, \partial_\mu T_j^{\mu \nu}
= S_j^\nu \enspace.
\end{equation}
The right-hand sides define the so-called {\em source terms\/} which
according to (\ref{ris:Nmuj}), (\ref{ris:Tmunuj}) obey
\begin{equation}
\sum_{j=1}^M S_{ij} = 0\,\,\, , \,\,\,\, \sum_{j=1}^M S_j^\nu = 0
\enspace .
\end{equation}
The source terms are parameters of a particular model and have to
be specified e.g.\ from kinetic theory. Let us consider the
Boltzmann equation for particles from fluid $j$:
\begin{equation}
k \cdot \partial \, f_j(k,x) = \sum_{klm} \left[{\cal C}_{lm}^{jk}
- {\cal C}_{jk}^{lm}\right] \enspace.
\end{equation}
The right-hand side involves the collision terms for the
microscopic 2-particle 
reactions $lm \rightarrow jk$ (the {\em gain term} ${\cal C}_{lm}^{jk}$) where
particles from fluid $l$ and fluid $m$ ($l$ and $m$ not necessarily
different) collide to produce particles of fluid $j$ and $k$
(again, $j$ and $k$ not necessarily different), and
$jk \rightarrow lm$ (the {\em loss term} ${\cal C}_{jk}^{lm}$) 
where particles from
fluid $j$ and $k$ collide to produce particles of fluid $l$ and $m$.
Taking the zeroth and first moment of this equation yields
\begin{eqnarray}
\partial \cdot N_{ij} & \equiv & q_i 
\int \frac{ \D^3 \vec{k}}{E}\, k \cdot \partial\, f_j(k,x) 
= q_i \sum_{klm} \int \frac{\D^3 \vec{k}}{E} \, 
\left[{\cal C}_{lm}^{jk} - {\cal C}_{jk}^{lm}\right] 
\equiv S_{ij} \enspace , \\
\partial_\mu T_i^{\mu \nu} & \equiv & 
\int \frac{ \D^3 \vec{k}}{E}\, k^\mu k^\nu\, \partial_\mu\, f_j(k,x) 
= \sum_{klm} \int \frac{\D^3 \vec{k}}{E} \, k^\nu
\left[{\cal C}_{lm}^{jk} - {\cal C}_{jk}^{lm}\right] 
\equiv S_{j}^\nu \enspace. 
\end{eqnarray}
This defines the source terms through the microscopic collision rates.

Results of any specific multi-fluid model will not be discussed here,
I instead refer the reader to the 
literature on this subject \cite{ris:multifluid}.
I close with two remarks: (a) a {\em single\/} fluid may consist
of several {\em different\/} particle species (for instance, $\pi,\, K, \,
N, \, \Lambda$ etc.), as long as it is reasonable to assume that they
stay in local thermodynamical equilibrium among each other. Then, the
only place where information enters about these different particle species
is the equation of state $p(\epsilon,n_1, \ldots, n_n)$.
(b) {\em Different\/} fluids may consist of the {\em same\/}
particle species (with the {\em same\/} equation of state $p(\epsilon,
n_1,\ldots,n_n)$). This situation occurs for instance in the initial stage
of relativistic heavy-ion collisions, where the single-particle 
phase space distributions of target and projectile nucleons, while
overlapping in space-time,
are still well separated in momentum space due to the high initial relative
velocity between them. This is a situation where there is 
local thermodynamical equilibrium in target and projectile separately, but
not between them. It therefore is reasonable to treat target and
projectile, although consisting of the same particle species, as two
separate fluids.

\section{Numerical Aspects}

In this section, I discuss basic aspects of
numerical solution schemes for relativistic ideal fluid dynamics.
For the sake of simplicity, let us consider 
the case of one conserved charge only. Define
\begin{eqnarray} \label{ris:R}
R & \equiv & N^0 = n\, u^0 = n\, \gamma \enspace, \\
E & \equiv & T^{00} = (\epsilon+p) \gamma^2 - p \enspace, 
\label{ris:E}\\
\vec{M} & \equiv & \left\{ T^{0i} \right\}_{i=x,y,z} = (\epsilon +p) \gamma^2
\vec{v} \enspace , \label{ris:M}
\end{eqnarray}
where $u^\mu = \gamma (1, \vec{v})$ is the fluid 4-velocity, $\gamma
= (1-\vec{v}^2)^{-1/2}$. With these definitions, the  
conservation laws (\ref{ris:localcons1}), (\ref{ris:localcons2}) take the form
\begin{eqnarray}
\partial \cdot N & \equiv & \partial_t R + \nabla \cdot (R \vec{v}) = 0
\enspace , \\
\partial_\mu \, T^{\mu 0} & \equiv & \partial_t E + \nabla \cdot
\left[ (E+p) \vec{v} \right] = 0 \enspace , \\
\partial_\mu \, T^{\mu i} & \equiv & \partial_t M^i + \nabla \cdot
(M^i \vec{v}) + \partial_i p = 0 \enspace .
\end{eqnarray}
In this form, the conservation equations can be solved numerically
with any scheme that also solves the non-relativistic conservation
equations. There is, however, one fundamental difference between the
non-relativistic equations and the relativistic ones.
In order to solve the latter
for $R,\, E,\, \vec{M}$, the net charge density, energy density, and
momentum density in the {\em calculational frame}, one has to
know the equation of state $p(\epsilon,n)$ and $\vec{v}$. The
equation of state, however, depends on $n,\, \epsilon$, 
the net charge density
and energy density {\em in the rest frame of the fluid}. One therefore
has to locally transform from the calculational frame to the rest frame
of the fluid in order to extract $n,\, \epsilon, \, \vec{v}$ from
$R,\, E,\, \vec{M}$. In the non-relativistic limit, there is no difference
between $n$ and $R$, or $\epsilon$ and $E$ and the equation of state can
be employed directly in the conservation equations. Also, the momentum
density of the fluid is related to the fluid velocity by a simple
expression. The transformation between rest frame and calculational frame
quantities is described explicitly in the next subsection.

\subsection{Transformation between Calculation Frame and Fluid Rest Frame}

In principle, the transformation is explicitly given by equations
(\ref{ris:R}) -- (\ref{ris:M}), i.e., by finding the roots of a set of 5
nonlinear equations (the non-linearity enters through the equation of
state $p(\epsilon,n)$). In numerical applications, however, this
transformation has to be done several times in each time step and each cell.
It is therefore advisable to reduce the complexity of the transformation
problem. This is done as follows \cite{ris:kaempfer}.

First note that $\vec{M}$ and $\vec{v}$ are parallel, thus
\begin{equation}
\vec{M} \cdot \vec{v} \equiv M\, v = (\epsilon+p) \gamma^2 v^2
= (\epsilon + p) (\gamma^2-1) = E- \epsilon \enspace ,
\end{equation}
where $M \equiv |\vec{M}|$, $v \equiv |\vec{v}|$.
Therefore,
\begin{equation} \label{ris:epsilon}
\epsilon = E- M\, v\,\,\, , \,\,\,\, n = R \sqrt{1-v^2}\enspace,
\end{equation}
where the second equation is a simple consequence of (\ref{ris:R}).
With these equations $\epsilon$ and $n$ can be expressed in terms of $R,\,
E,\, M$ and $v$. The 5-dimensional root search is therefore reduced
to finding the modulus of $v$ for given $R,\, E,$ and $M$, which is a 
simple one-dimensional problem. To solve this, use the definition of
$M$, 
\begin{equation}
M= (\epsilon+p) \gamma^2 v = (E+p) v \enspace .
\end{equation} 
This equation can be rewritten as a fixed point equation for $v$
for given $R,\, E,\, M$:
\begin{equation}
v = \frac{M}{E+p \left(E-M\, v, R \sqrt{1-v^2}\right)} \enspace.
\end{equation}
The fixed point yields the modulus of the fluid velocity, from which
one can reconstruct $\vec{v} = v \, \vec{M}/M$, and find $\epsilon$ and
$n$ via (\ref{ris:epsilon}). The equation of state $p(\epsilon, n)$ then
yields the final unknown variable, the pressure $p$.

\subsection{Operator Splitting Method}

In general, to model a heavy-ion collision with ideal fluid dynamics
requires to solve the 5 conservation equations in three space dimensions.
Since this is in general a formidable numerical task, one usually resorts 
to the so-called {\em operator splitting method\/}, i.e., the
full 3-dimensional solution is constructed by solving sequentially 
three one-dimensional problems. More explicitly, all conservation
equations are of the type
\begin{equation} \label{ris:dU}
\partial_t \, U + \sum_{i = x,y,z} \partial_i F_i(U) = 0 \enspace ,
\end{equation}
$U$ being $R,\, E,$ or $M^i$.
Such an equation is numerically solved on a space-time grid, 
and time and space derivatives are replaced by finite differences:
\begin{equation} \label{ris:dUfindiff}
U_{ijk}^{n+1} = U_{ijk}^{n} - \Delta t \, G\left[U_{ijk}^n\right] \enspace ,
\end{equation}
where $i,\,j,\,k$ are cell indices (the cell number 
in $x,\, y,$ and $z$ direction) and $n$ denotes the time step. 
$\Delta t$ is the time step width.
$G\left[U_{ijk}^n\right]$ is a suitable finite difference form of
the 3-divergence in (\ref{ris:dU}).

It can be shown that in the continuum limit instead of
solving (\ref{ris:dUfindiff}) it is equivalent to
solve the following set of {\em predictor-corrector\/} equations
\begin{eqnarray}
U_{ijk}^{(1)\, n+1} & = & U_{ijk}^n - \Delta t \, G_x \left[U_{ijk}^n \right]
\enspace , \nonumber \\
U_{ijk}^{(2)\, n+1} & = & U_{ijk}^{(1) \,n+1} - \Delta t \, 
G_y \left[U_{ijk}^{(1)\, n+1} \right] \enspace , \\ 
U_{ijk}^{n+1} & = & U_{ijk}^{(2)\,n+1} - \Delta t \, 
G_z \left[U_{ijk}^{(2)\, n+1} \right] \enspace , 
\end{eqnarray}
and that the solution converges towards the solution of (\ref{ris:dU}).
Here, the $G_i [U],$ $i=x,y,z,$ are finite difference forms of
the partial derivatives $\partial_i \, F_i(U)$ (no summation over $i$)
in $x,\, y,$ or $z$ direction.
$U_{ijk}^{(1)\, n+1}$ is the first {\em prediction\/} for the full solution
$U_{ijk}^{n+1}$. It is generated by solving a finite difference form
of the one-dimensional equation
\begin{equation} \label{ris:dUi}
\partial_t \,U + \partial_i\, F_i(U) = 0 \enspace ,
\end{equation}
where $i=x$. Subsequently, the first prediction $U_{ijk}^{(1)\, n+1}$ is 
used to solve a finite difference form of (\ref{ris:dUi}), where now
$i=y$, to obtain the second {\em prediction\/} $U_{ijk}^{(2)\, n+1}$ for
the full solution. ($U_{ijk}^{(1)\, n+1}$ has been {\em corrected\/}
to $U_{ijk}^{(2)\, n+1}$.)
Finally, the full solution $U_{ijk}^{n+1}$ is obtained by using
$U_{ijk}^{(2)\, n+1}$ to solve a finite difference form of (\ref{ris:dUi})
with $i=z$. ($U_{ijk}^{(2)\, n+1}$ has been {\em corrected\/}
to $U_{ijk}^{n+1}$.)

In other words, the solution to the partial differential equation
(\ref{ris:dU}) in three space dimensions is obtained by solving
a sequence of partial differential equations (\ref{ris:dUi})
in one space dimension. 
The 3-divergence operator in (\ref{ris:dU}) was {\em split\/} into
a sequence of three partial derivative operators.
Physically speaking, in a given time step
one first propagates the fields in $x$ direction, then in $y$ direction, and
then in $z$ direction. In actual numerical applications it is
advisable to permutate the order $xyz$ to minimize systematical errors.

The advantage of the operator splitting method is that there exists a
variety of numerical algorithms which solve evolution equations 
of the type (\ref{ris:dUi}) in one space dimension (see, for instance,
\cite{ris:schneider} and refs.\ therein). 
One of them is discussed in the following subsection.

\subsection{The Relativistic Harten--Lax--van Leer--Einfeldt Algorithm}

The relativistic Harten--Lax--van Leer--Einfeldt (HLLE) algorithm
\cite{ris:schneider,ris:RHLLE} solves equations of the type 
\begin{equation} \label{ris:dU2}
\partial_t \, U + \partial_x\, F(U) = 0 \enspace ,
\end{equation}
i.e., propagation of a field $U$ in one space dimension.
For ideal relativistic fluid dynamics, $U = R,\, E,$ or $M$ and
$F(U) = Rv,\, (E+p)v$, or $Mv+p$. (For one-dimensional propagation, it
is sufficient to consider only the components of the momentum density
$\vec{M}$ and the fluid velocity $\vec{v}$ in the direction
of propagation. They are here denoted by $M$ and $v$, respectively.)

\begin{figure}
\psfig{file=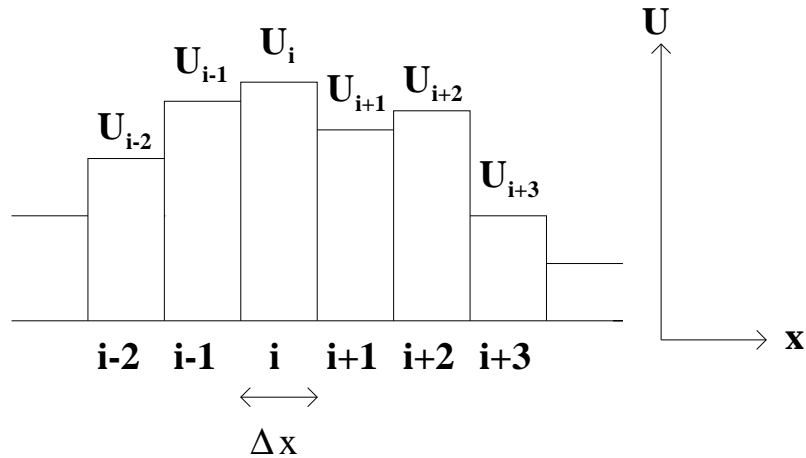}
\caption{The initial distribution of the density $U$ on the
numerical grid.} \label{ris:Fig1}
\end{figure}

The idea behind the relativistic HLLE scheme is the following.
Consider the initial distribution of the density 
$U$ on a numerical grid. $U$ is assumed to be constant inside each cell,
but different from cell to cell, i.e.,
the initial distribution consists of a sequence of constant flow fields
inside the cells separated by discontinuities at the cell boundaries, 
cf.\ Fig.\ \ref{ris:Fig1}.

\begin{figure}
\hspace*{2cm}
\psfig{file=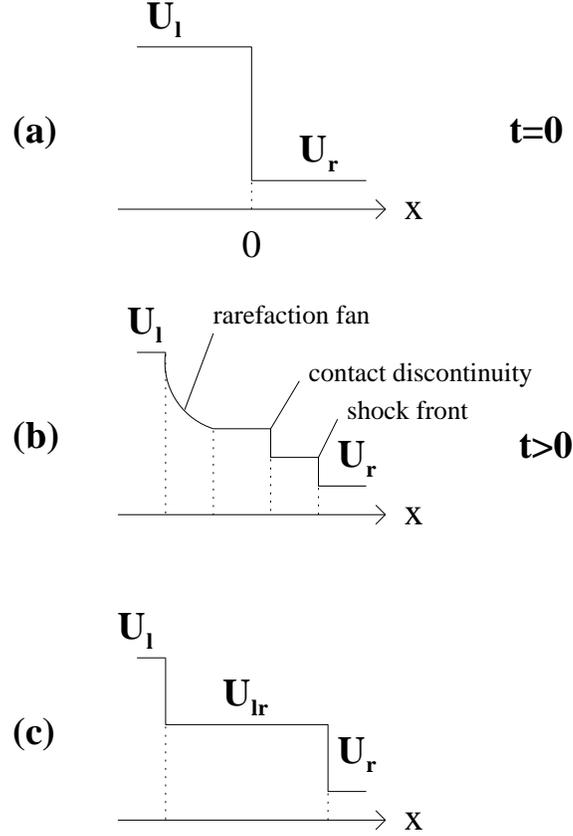}
\caption{(a) The initial condition of the Riemann problem at $t=0$.
(b) The solution of the Riemann problem at $t>0$.
(c) The approximate solution of a Godunov-type algorithm.} \label{ris:Fig2}
\end{figure}

In the further time evolution these discontinuities will decay, resulting
in the transport of $U$ across the grid.
The decay of a discontinuity between two regions of constant flow
is, however, a well-known problem in fluid dynamics, the so-called
{\em Riemann problem}. For simple equations of state it is even analytically
solvable. Consider the discontinuity to be located at $x=0$.
Denote the density in the region of
constant flow to the left of the discontinuity by $U_{\rm l}$, and that to the
right by $U_{\rm r}$. The initial condition at time $t=0$ then reads
\begin{equation}
U(x,0) = \left\{ \begin{array}{ll}
                 U_{\rm l} \,\,, & \,\,  x<0 \\
                 U_{\rm r} \,, & \,\, x \geq 0 
                 \end{array} \right. \enspace,
\end{equation}
cf.\ Fig.\ \ref{ris:Fig2} (a). For the sake of definiteness, consider
$U_{\rm l} > U_{\rm r}$. For $t>0$, the solution looks qualitatively
as in Fig.\ \ref{ris:Fig2} (b). There is a rarefaction fan propagating
into the region of higher density (in this case to the left), and a 
shock front into the region of lower density (in this case to the right).
Between fan and shock wave there are two regions of constant flow separated
by a contact discontinuity (a discontinuity where the pressure is equal
on both sides). It is evident that a numerical algorithm can be constructed
which solves the fluid dynamical equations simply by solving a sequence 
of Riemann problems for the discontinuities at all cell boundaries in 
a given time step. Such algorithms 
are called {\em Godunov algorithms\/} \cite{ris:holt}.

The relativistic HLLE is a so-called {\em Godunov-type\/} algorithm
\cite{ris:holt}, i.e., it does not employ the full solution of the Riemann 
problem but approximates it by a region of constant flow between $U_l$ and
$U_r$, cf.\ Fig.\ \ref{ris:Fig2} (c):
\begin{equation}
U(x,t) = \left\{ \begin{array}{ll}
                 U_{\rm l} \,\, ,  & \,\, x < b_{\rm l} \, t \\
                 U_{\rm lr} \,\, , & \,\, b_{\rm l}\, t \leq x < b_{\rm r} 
                                                                 \, t \\
                 U_{\rm r} \,\, , & \,\, x \geq b_{\rm r}\, t 
                 \end{array} \right. \enspace .
\end{equation}
Here, $b_{\rm l} < 0$ and $b_{\rm r} >0$ are the so-called 
{\em signal velocities}.
They characterize the velocities with which information about the
decay of the discontinuity travels to the left and right into the regions
of constant flow. The value $U_{\rm lr}$ in the region of constant flow
between $U_{\rm l}$ and $U_{\rm r}$ is determined in accordance with the 
conservation laws. To this end, integrate (\ref{ris:dU2}) over a fixed 
interval $[x_{\rm min},x_{\rm max}],\, x_{\rm min} < b_{\rm l}\, t,\, 
x_{\rm max} > b_{\rm r} \, t$.
One obtains:
\begin{equation} \label{ris:Ulr}
U_{\rm lr} = \frac{b_{\rm r} \, U_{\rm r} - b_{\rm l}\, U_{\rm l} + 
F(U_{\rm l}) - F(U_{\rm r})}{b_{\rm r} - b_{\rm l}}
\enspace .
\end{equation}
The value of the flux $F(U_{\rm lr})$ corresponding to the density 
$U_{\rm lr}$ is
determined by integrating (\ref{ris:dU2}) over the fixed interval
$[0,x_{\rm max}]$ or $[x_{\rm min},0]$:
\begin{equation} \label{ris:FUlr}
F(U_{\rm lr}) = \frac{b_{\rm r}\, F(U_{\rm l}) - b_{\rm l}\, F(U_{\rm r}) 
+ b_{\rm l}\, b_{\rm r}\, (U_{\rm r}-U_{\rm l})}{b_{\rm r} - b_{\rm l}} 
\enspace .
\end{equation}
Upon discretization, the differential operator $\partial_x\, F(U)$ in
the evolution equation for the density $U_i$ in cell $i$ assumes
the form $[F(U_{i+1/2}) - F(U_{i-1/2})]/\Delta x$ where $\Delta x$
is the cell size (grid spacing) and $U_{i \pm 1/2}$ are values of
the density at the position of the right and left boundary of cell $i$. 
These values are taken {\em after\/} the decay of the respective
discontinuities at the cell boundaries, i.e., they are the corresponding
values $U_{\rm lr}$ given by (\ref{ris:Ulr}) and the respective
$F(U_{i \pm 1/2})$ are the corresponding values $F(U_{\rm lr})$ given
by (\ref{ris:FUlr}). This yields the following explicit expressions for
the relativistic HLLE scheme
\begin{eqnarray}
U_i^{n+1} & = & U_i^n - \frac{\Delta t}{\Delta x} \left[
F\left(U_{i+1/2}^n\right) - F\left(U_{i-1/2}^n \right) \right] \enspace , \\
F\left(U_{i+1/2}^n\right) & = & \frac{b_{\rm r}\, F\left(U_i^n\right)
- b_{\rm l}\, F\left(U_{i+1}^n\right) + b_{\rm r}\,b_{\rm l}
\left(U_{i+1}^n - U_i^n\right) }{b_{\rm r} - b_{\rm l}} \enspace .
\end{eqnarray}
A reasonable estimate for the signal velocities is to take them as 
the relativistic addition (subtraction) of flow velocities and 
sound velocities in the respective cells adjacent to the cell boundary:
\begin{eqnarray}
b_{\rm r} & = &\max \left\{ 0, \frac{v_{i+1}^n + c_{{\rm s},i+1}^n}{1
+v_{i+1}^n\, c_{{\rm s},i+1}^n} \right\} \enspace , \\
b_{\rm l} & = &\min \left\{ 0, \frac{v_{i}^n - c_{{\rm s},i}^n}{1-v_{i}^n\, 
c_{{\rm s},i}^n} \right\} \enspace .
\end{eqnarray}
As described above, this scheme is accurate to first order in time. A
scheme which is accurate to second order can be obtained using half-step
updated values $F\left(U_{i\pm 1/2}^{n+1/2}\right)$, 
for more details see \cite{ris:dhr}.

\section{One-dimensional Solutions}

In this section I discuss solutions of ideal relativistic fluid dynamics
in one space dimension. I first introduce the notion of characteristic curves.
Then, I discuss possible one-dimensional wave patterns 
for thermodynamically normal and anomalous media. Choosing a representative
equation of state which features both thermodynamically normal and anomalous
regions I then discuss the expansion of semi-infinite matter into the
vacuum. The emerging wave patterns will help us to understand the possible
solutions of the Landau model, which was historically the first fluid-dynamical
model for relativistic heavy-ion collisions. Finally, also the
Bjorken model for ultrarelativistic heavy-ion collisions is discussed.

\subsection{One-dimensional Wave Patterns}

\begin{figure}
\hspace*{1.5cm}
\psfig{file=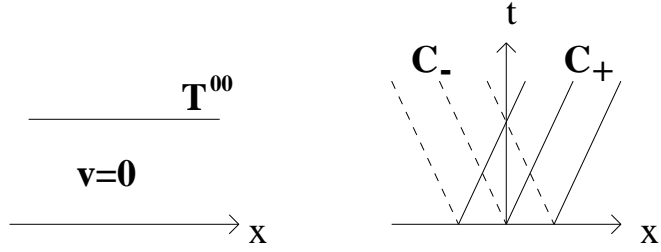}
\caption{The characteristic curves for a constant flow pattern.} 
\label{ris:Fig3}
\end{figure}

For flow in one spatial dimension (say, in $x$ direction)
the two conservation equations for energy and for
momentum read:
\begin{equation} \label{ris:1dcons}
\partial_t\, T^{00} + \partial_x \, T^{x0} = 0 \,\,\,\, , \,\,\, \,\,
\partial_t\, T^{0x} + \partial_x \, T^{xx} = 0 \enspace .
\end{equation}
A suitable linear combination of these equations leads to
the equivalent set of equations
\begin{equation} \label{ris:charact}
\left( \partial_t + \frac{v\pm c_{\rm s}}{1 \pm v\, c_{\rm s}} \,
\partial_x \right) {\cal R}_{\pm} = 0 \enspace ,
\end{equation}
where $c_{\rm s}^2 \equiv \partial p /\partial \epsilon |_{s/n}$ is
the velocity of sound squared ($s/n$ is the specific entropy) and
\begin{equation} \label{ris:Riemann}
{\cal R}_{\pm} \equiv y - y_0 \pm \int_{\epsilon_0}^\epsilon
\D \epsilon' \, \frac{c_{\rm s}(\epsilon')}{\epsilon' + p(\epsilon')}
\end{equation}
are the so-called {\em Riemann invariants}, $y \equiv {\rm Artanh} v$ is
the fluid rapidity. Equation (\ref{ris:charact}) has the obvious
interpretation that the Riemann invariants
${\cal R}_{\pm}$ are constant along world lines $x_\pm(t)$ defined by
\begin{equation} \label{ris:1charact}
\frac{\D x_{\pm}(t)}{\D t} \equiv w_\pm = \frac{v \pm c_{\rm s}}{1
\pm v\, c_{\rm s}} \enspace .
\end{equation}
These world lines are the so-called {\em characteristic curves\/}
or {\em characteristics} ${\cal C}_\pm(x,t)$. 
It is also obvious that these curves are
the world lines of {\em sonic perturbations\/} or {\em sound waves\/}
on top of the fluid-dynamical wave pattern. ${\cal C}_+(x,t)$ characterizes
sound waves moving to the right (in positive $x$ direction) while
${\cal C}_-(x,t)$ characterizes those moving to the left (in negative
$x$ direction). For the simple example of constant flow, the characteristic
curves are shown in Fig.\ \ref{ris:Fig3}.

\begin{figure}
\hspace*{3cm}
\psfig{file=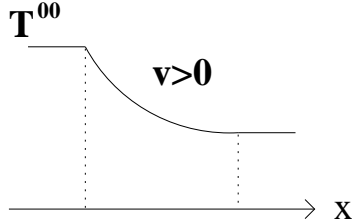}
\caption{A continuous simple wave between
two regions of constant flow, moving to the right.}
\label{ris:Fig4}
\end{figure}

Let us now consider a so-called {\em simple rarefaction wave\/} 
moving to the right, cf.\ Fig.\ \ref{ris:Fig4}.
(For the definition of a simple wave, see \cite{ris:courant}, for our
purposes it is sufficient to remark that in one spatial dimension
a simple wave is the only possible wave that can connect two regions
of constant flow. A rarefaction wave denotes a wave where the energy density
decreases in the direction of propagation.) 
Then, one can prove that ${\cal R}_+ = {\rm const.}$
everywhere (for the proof, see \cite{ris:courant}; analogously, for
simple waves moving to the left, ${\cal R}_- = {\rm const.}$).

\begin{figure}
\psfig{file=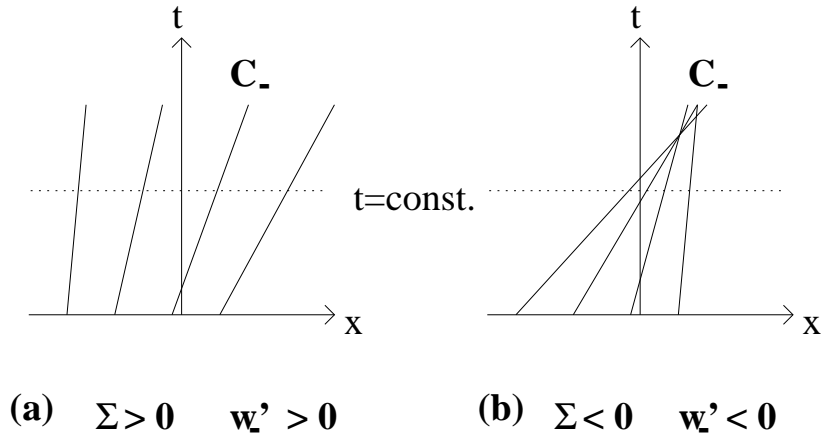}
\caption{For a simple wave moving to the right and (a) $\Sigma >0$
the ${\cal C}_-$ characteristics fan out, while for (b) $\Sigma <0$
they converge and intersect.}
\label{ris:Fig5}
\end{figure}

It is therefore sufficient to consider the equation for the
${\cal R}_-$ invariants, or the ${\cal C}_-$ characteristics, respectively.
Let us consider how the slope of the ${\cal C}_-$ characteristics
changes with $x$ at constant $t$:
\begin{equation}
\left. \frac{\partial \, w_-}{\partial x} \right|_t \equiv w_-' 
= \frac{v' (1- c_{\rm s}^2) - c_{\rm s}' (1-v^2)}{(1-v\, c_{\rm s})^2}
\enspace .
\end{equation}
From ${\cal R}_+ = {\rm const.}$ everywhere one infers
\begin{equation}
v' = - (1-v^2) \, \frac{c_{\rm s}}{\epsilon + p} \, \epsilon' \enspace ,
\end{equation}
while
\begin{equation}
c_{\rm s}' = \frac{1}{2 \, c_{\rm s}}\, \left. \frac{\partial^2 p}{\partial
\epsilon^2} \right|_{s/n} \epsilon' \enspace .
\end{equation}
Therefore,
\begin{equation} \label{ris:wminusprime}
w_-' = - \frac{1-w_-^2}{2\, c_{\rm s} (1- c_{\rm s}^2)} \, \Sigma \,
\epsilon' \enspace ,
\end{equation}
where
\begin{equation} \label{ris:Sigma}
\Sigma \equiv \left. \frac{\partial^2 p}{\partial \epsilon^2} \right|_{s/n}
+ 2\, c_{\rm s}^2 \, \frac{1-c_{\rm s}^2}{\epsilon +p} \enspace .
\end{equation}
Equation (\ref{ris:wminusprime}) is an important qualitative result:
Since the first factor is always positive ($w_-$ as well as $c_{\rm s}$
are causal), and since the energy density decreases with $x$ for the
rarefaction wave considered here, $\epsilon' <0$, the sign of $w_-'$ is solely
determined by the sign of $\Sigma$. The quantity $\Sigma$, however,
is solely determined by the equation of state of matter under consideration,
i.e., its sign (and absolute value) is an intrinsic property of the fluid. 
Matter with $\Sigma >0$ is called {\em thermodynamically normal}, while
matter with $\Sigma <0$ is {\em thermodynamically anomalous}.
More specifically,
if $\Sigma >0$, then $w_-' >0$, and if $\Sigma <0$, then $w_-' <0$.
A positive $w_-'$, however, means that the ${\cal C}_-$ characteristics
``fan out'' in the $x-t$ plane, while a negative $w_-'$ indicates that
they converge and ultimately intersect at one point, 
cf.\ Fig.\ \ref{ris:Fig5}.

Intersecting characteristics, however, signal the formation of
{\em shock waves}. Physically speaking, picture a sonic
perturbation (travelling along a characteristic) emitted at
a point $x_1$. This perturbation will eventually overtake a 
perturbation emitted at $x_2 > x_1$ (namely when the corresponding
characteristics intersect). Thus,
the two small perturbations add up to form a larger one. Imagine this 
happening for other perturbations (emitted at different points) as well.
Eventually, a finite discontinuity (shock wave) is formed from the 
superposition of a large number of infinitesimal sonic perturbations.
Shock waves are discontinuous solutions of ideal fluid dynamics and
will be discussed in more detail in the following subsection.

I conclude this subsection by collecting the above arguments in 
the following classification scheme of one-dimensional wave patterns.
{\em Continuous rare\-faction waves\/} are {\em stable\/} 
in {\em thermodynamically normal matter\/} 
while they are {\em unstable\/} in {\em anomalous matter}.
On the other hand, {\em rarefaction shock waves\/} are {\em stable\/} in
{\em thermodynamically anomalous matter\/} while they are 
{\em unstable\/} in {\em thermodynamically normal matter}.
If we perform an analogous consideration for a {\em continuous 
compression wave\/} we are led to the conclusion that
such waves are {\em unstable\/} in {\em normal\/} and {\em stable\/} 
in {\em anomalous\/} matter, while {\em compression shock waves\/} are
{\em stable\/} in {\em normal\/} and {\em unstable\/} in
{\em anomalous\/} matter.
These results are summarized in Table \ref{ris:table1}. A ``$+$'' sign
means ``stable'' while a ``$-$'' sign indicates ``unstable''.

\begin{table} \label{ris:table1}
\caption{Classification scheme for the 
stability of one-dimensional wave patterns.}
\renewcommand{\arraystretch}{1.2}
\begin{tabular}{l|c|c}
\hline\noalign{\smallskip}
Wave                   & $\Sigma>0$  & $\Sigma<0 $  \\
\noalign{\smallskip}
\hline
\noalign{\smallskip}
Continuous rarefaction &     $+$    &    $-$ \\
Rarefaction shock      &     $-$    &    $+$ \\ 
Continuous compression &     $-$    &    $+$ \\
Compression shock      &     $+$    &    $-$ \\
\noalign{\smallskip}
\hline
\end{tabular}
\renewcommand{\arraystretch}{1.0}
\end{table}

Most matter is thermodynamically normal. In the presence of phase transitions,
however, an equation of state can feature regions where matter 
is thermodynamically anomalous. As will be seen in Subsections 4.4 and 4.5, 
this will strongly influence the time evolution of the system
in a qualitative and quantitative way.

\subsection{Shock discontinuities}

\begin{figure}
\hspace*{3cm}
\psfig{file=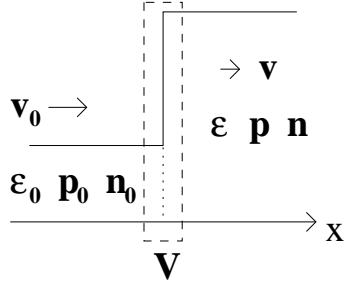}
\caption{A shock discontinuity in its rest frame.}
\label{ris:Fig6}
\end{figure}

Shock waves represent discontinuous solutions of ideal fluid dynamics.
While the partial derivatives of $N_i^\mu$ and $T^{\mu \nu}$ appearing in the
conservation equations are ill-defined at the location of such
discontinuities, there is still a simple way solve the problem of
charge and energy-momentum transport across a shock discontinuity.
To this end, let us consider the case of one conserved charge only,
and study such a discontinuity in its rest frame.
Matter enters the discontinuity with velocity $v_0$ in a thermodynamic
state characterized by the net charge density $n_0$, the energy density
$\epsilon_0$, and the pressure $p_0$ (which is of course determined by
$\epsilon_0$ and $n_0$ through the equation of state). The task is to
determine the velocity $v$ and the thermodynamic state of matter
($n,\, \epsilon$, and $p$) emerging from the shock.
Imagine a small volume $V$ which encloses the discontinuity, cf.\ Fig.\
\ref{ris:Fig6}.

Let us now integrate the conservation equations (\ref{ris:localcons1}) 
(for a single conserved charge) and
(\ref{ris:localcons2}) for one-dimensional flow over $V$:
\begin{eqnarray}
\partial_t \int_V \D^3 \vec{x} \, N^{0} + \int_V \D^3 \vec{x}\,
\partial_x\, N^{x} & = & 0 \enspace , \\
\partial_t \int_V \D^3 \vec{x} \, T^{00} + \int_V \D^3 \vec{x}\,
\partial_x\, T^{x0} & = & 0 \enspace , \\
\partial_t \int_V \D^3 \vec{x} \, T^{x0} + \int_V \D^3 \vec{x}\,
\partial_x\, T^{xx} & = & 0 \enspace .
\end{eqnarray}
In a steady-state situation (a stable, propagating shock discontinuity)
the total amount of charge, energy and momentum inside $V$ cannot
change with time, therefore, the first terms in these equations vanish.
The other terms are integrated by parts to yield the set of equations
\begin{eqnarray} \label{ris:ngv}
n_0 \gamma_0 v_0 & = & n\, \gamma \, v \enspace , \\
(\epsilon_0+p_0) \gamma_0^2 v_0 & = & (\epsilon +p)\, \gamma^2 \, v 
\enspace , \\
(\epsilon_0+p_0) \gamma_0^2 v_0^2 + p_0 & = & (\epsilon +p)\, \gamma^2 \, v^2 
+ p \enspace . \label{ris:epg2}
\end{eqnarray}
These are the conservation equations for net charge and
energy-momentum across a shock discontinuity. They are no longer
partial differential equations, but purely algebraic. For a 
given initial state $n_0,\,\epsilon_0,\, p_0$, and velocity $v_0$, they
determine the final state $n,\, \epsilon,\, p$, and the velocity $v$
of compressed matter emerging from the shock, if the equation of state
$p(\epsilon,n)$ is known.

\begin{figure}
\psfig{file=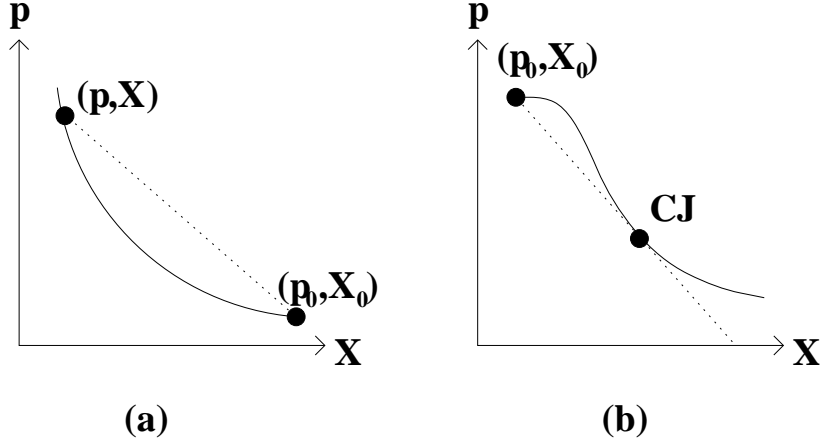}
\caption{(a) The Taub adiabat for a compressional shock wave.
$(p_0,X_0)$ is the center of the adiabat, $(p,X)$ is one final
state on the adiabat which is selected by a choice of $v_0$.
(b) The Taub adiabat for a rarefaction shock wave. ``CJ''
denotes the Chapman--Jouguet point.}
\label{ris:Fig7}
\end{figure}

One can eliminate the velocities from the set of equations (\ref{ris:ngv})
-- (\ref{ris:epg2}) to obtain the so-called {\em Taub equation\/} 
\cite{ris:taub}
\begin{equation}
(\epsilon +p) X - (\epsilon_0 +p_0) X_0 =
(p-p_0) ( X+X_0 ) \enspace ,
\end{equation}
where $X\equiv (\epsilon+p)/n^2$ is the so-called {\em generalized
volume}. Once $p(\epsilon,n)$ is fixed, the solution of the Taub
equation defines the so-called {\em Taub adiabat\/} $p(X)$,
cf.\ Fig.\ \ref{ris:Fig7}. For a given initial state $(p_0,X_0)$
(the so-called {\em center\/} of the adiabat) it represents all 
final states $(p,X)$ for matter emerging from the shock, 
which are in agreement with net charge and energy-momentum conservation. 
The actual final state is then selected by specifying $v_0$.
This determines all variables uniquely in the rest frame of the shock.
The remaining unknown is, however, the velocity of the shock in an
arbitrary calculational frame. For compressional shock waves, such as occur in
the initial stage of heavy-ion collisions (cf.\ \cite{ris:dhr2} for a
detailed discussion), this shock velocity can be uniquely
determined from the geometry of the collision. For rarefaction shock
waves this is not possible, and thus in principle there is a whole
region of final states on the Taub adiabat, which are in agreement
with energy-momentum and net charge conservation. It turns out, however,
that the stationary situation is always given by a rarefaction shock 
where matter emerges at the so-called {\em Chapman--Jouguet\/} point,
indicated by ``CJ'' in Fig.\ \ref{ris:Fig7} (b) \cite{ris:LL}. This point
is defined as the point where a chord between the center $(p_0,X_0)$
and a final state on the adiabat is tangential to the adiabat.
This then uniquely fixes the state of matter emerging from the shock, 
as well as the velocity of the shock in the calculational frame.
Note that it is also possible to define a Taub adiabat in the case that
there is no conserved charge, see \cite{ris:dhr,ris:ruuskanen} for details.

To conclude this subsection, let us consider what happens to the entropy
flux across a shock discontinuity. Integrate the conservation equation
(\ref{ris:entropycons}) for the entropy current over the volume $V$ which
encloses the shock front in its rest frame,
\begin{equation}
\partial_t \int_V \D^3 \vec{x}\, s\, \gamma + \int_V \D^3\vec{x} \, \partial_x
\, s\, \gamma\, v = 0 \enspace ,
\end{equation}
and perform an integration by parts in the second term. This yields:
\begin{equation}
s\, \gamma\, v = s_0 \gamma_0 v_0 + \frac{1}{A_\perp} \, \partial_t \, S
\enspace ,
\end{equation}
where $A_\perp$ is the transverse area of the shock front and
$S \equiv \int_V \D^3 x \, s\, \gamma$ is the total entropy inside the
volume $V$. The second law of thermodynamics tells us that the entropy cannot
decrease,  $\partial_t \, S \geq 0$. Consequently,
\begin{equation}
s\, \gamma\, v \geq s_0 \gamma_0 v_0 \enspace .
\end{equation}
Dividing both sides by (\ref{ris:ngv}) one concludes
\begin{equation} \label{ris:sovern}
\frac{s}{n} \geq \frac{s_0}{n_0}  \enspace ,
\end{equation}
i.e., the specific entropy increases across a shock front. 
This result is remarkable, since we know that the entropy current is
conserved in ideal fluid dynamics, Eq.\ (\ref{ris:entropycons}). However,
this equation holds strictly only for continuous (differentiable) solutions.
Shock discontinuities do not belong to this class, and therefore can
produce entropy. Physically speaking, microscopic non-equilibrium
processes take place inside a shock front which lead to this increase
of entropy. 

One could object that this conclusion is not stringent in the sense that
(\ref{ris:sovern}) also allows for the case where $s/n = s_0/n_0$, i.e., 
where the entropy does not increase across the shock front. However, by
explicitly solving the shock equations (\ref{ris:ngv}) -- (\ref{ris:epg2})
with a given equation of state one finds that this case occurs only
for infinitesimal shock discontinuities (which then degenerate into
sonic perturbations, which in turn preserve entropy). For any finite
discontinuity one finds $s/n > s_0/n_0$. 

The Chapman--Jouguet point 
(cf.\ Fig.\ \ref{ris:Fig7}) is actually special in this respect: it  
corresponds to that state of matter emerging from the shock, where 
entropy production is maximized \cite{ris:LL}. It is amusing to note
that in selecting this state as the final state of matter emerging 
from a rarefaction shock wave (cf.\ discussion above), fluid dynamics
not only automatically respects the second law of thermodynamics, but
even exploits it to the maximum extent.

\subsection{Equation of State and Expansion into Vacuum}

In this subsection I discuss possible wave patterns for the one-dimensional
expansion of semi-infinite matter into the vacuum. To be specific, let us
first choose an equation of state which bears relevance to relativistic
heavy-ion physics. At zero net baryon number,
QCD lattice data \cite{ris:laermann} suggest the following
Ansatz for the entropy density as function of temperature:
\begin{equation}
s(T) = c_{\rm H} T^3 \frac{1-\tanh [ (T-T_{\rm c})/\Delta T ] }{2}
     + c_{\rm Q} T^3 \frac{1+\tanh [ (T-T_{\rm c})/\Delta T ] }{2} \enspace ,
\end{equation}
where $c_{\rm Q}/c_{\rm H}$ is the ratio of degrees of freedom in
the quark-gluon phase and the hadronic phase, $T_{\rm c} \simeq 160$ MeV
is the (phase) transition temperature, and $\Delta T $
is the width of the transition. Present lattice data are not yet
sufficiently precise to decide whether the transition is first 
(corresponding to $\Delta T = 0$) or higher order, or just a smooth
cross-over transition, but they restrict $\Delta T$ to be within the range
$0 \leq \Delta T \la 0.1\, T_{\rm c}$. 
Note that for $\Delta T=0$ the equation 
of state becomes that of the well-known MIT bag model \cite{ris:MIT}
with a bag constant $B=(c_{\rm Q}/c_{\rm H} -1) p_{\rm c}$, 
where $p_{\rm c}$ is the pressure at the phase transition 
temperature $T_{\rm c}$. 

\begin{figure}
\hspace*{0.3cm}
\psfig{file=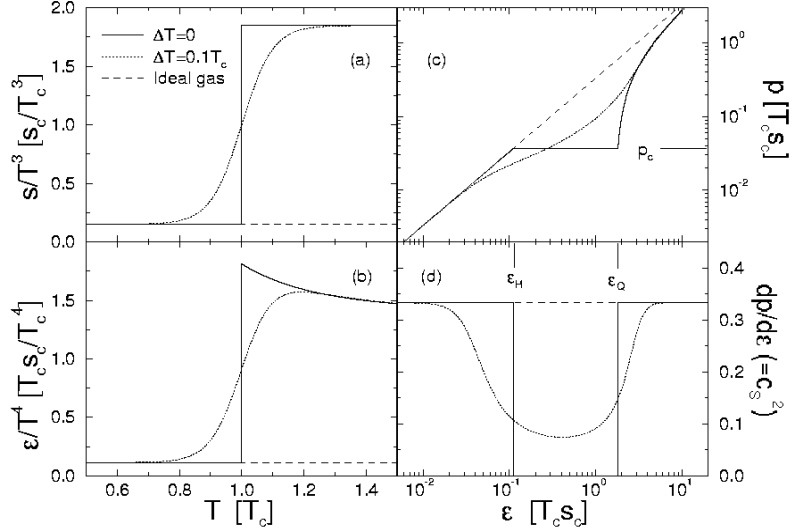,width=4.2in,height=3in,angle=-90}
\caption{(a) The entropy density divided by $T^3$ as a function of
$T$. (b) The energy density divided by $T^4$ as a function of $T$.
(c) The pressure as a function of energy density.
(d) The velocity of sound squared as a function of energy density.
$c_{\rm Q}/c_{\rm H} = 37/3$. Units of energy are $T_{\rm c}$, units of
energy density are $T_{\rm c} s_{\rm c}$, 
where $s_{\rm c} \equiv (c_{\rm Q} + c_{\rm H})\, T_{\rm c}^3/2$. Solid line:
$\Delta T=0$, dotted line: $\Delta T = 0.1\, T_{\rm c}$, 
dashed line: ideal hadron gas.}
\label{ris:Fig8}
\end{figure}

To cover the possible range of $\Delta T$, 
we shall consider the limiting values 
$\Delta T=0$ and $\Delta T = 0.1\, T_{\rm c}$ in the following. 
Both cases will be compared to results for an equation of state where 
there is no transition to the quark-gluon phase, i.e., where
\begin{equation}
s(T) \equiv s_{\rm H} (T) = c_{\rm H} T^3 \enspace .
\end{equation}
Once $s(T)$ is known one can compute other thermodynamic variables
from fundamental thermodynamic relations, for instance:
\begin{equation}
p = \int_0^T \D T'\, s(T') \,\,\, ,\,\,\,\, \epsilon  = T\, s - p \enspace .
\end{equation}
The three equations of state considered here are explicitly shown in
Fig.\ \ref{ris:Fig8}. The ratio of degrees of freedom 
$c_{\rm Q}/c_{\rm H}$ was chosen to be 37/3, corresponding to an 
ultrarelativistic gas of $u$ and $d$ quarks and gluons in the 
quark-gluon phase and a massless pion gas in the hadronic phase. 

Figs.\ \ref{ris:Fig8} (a,b) show the entropy density divided by $T^3$
and the energy density divided by $T^4$ as functions of $T$. This
representation of the equation of state is commonly used by the lattice
QCD community. On the other hand, fluid dynamics requires the
pressure as a function of energy density, $p(\epsilon)$,
which is shown in Fig.\ \ref{ris:Fig8} (c). The collective evolution
of the fluid is, however, controlled by pressure {\em gradients}.
Figure \ref{ris:Fig8} (d) shows the velocity of sound squared
$c_{\rm s}^2 \equiv \D p/\D \epsilon$ (if there are no conserved charges).
This quantity determines the pressure gradient $\D p$ for a given
gradient in energy density $\D \epsilon$, i.e., it characterizes the
capability of the fluid to perform mechanical work, or in other words,
it characterizes the {\em expansion tendency}. Thus, for the equation of
state with a first order phase transition, $\Delta T=0$, in the mixed phase
of quark-gluon and hadronic matter, $\epsilon_{\rm H} \leq \epsilon
\leq \epsilon_{\rm Q}$, the system does {\em not\/} perform
mechanical work and therefore has {\em no\/} tendency to expand.
As will be seen in the following this will have profound 
influence on the time evolution of the system.

For the equation of state with a smooth cross-over transition,
$\Delta T = 0.1\,T_{\rm c}$, the expansion tendency is not zero, but
still greatly reduced in the transition region
as compared to the ideal gas equation of state 
without any transition ($c_{\rm s}^2 = 1/3 = {\rm const.}$ for all
values of $\epsilon$). The transition region $\epsilon_{\rm H} \la
\epsilon \la \epsilon_{\rm Q}$ is referred to as the ``soft region''
of the equation of state \cite{ris:dhrmg}. For an
equation of state with a first order transition, the point $\epsilon =
\epsilon_{\rm Q}$ is called the ``softest point'' of the equation
of state \cite{ris:hung}. (This notion comes from considering the
function $p(\epsilon)/\epsilon$ which has a minimum 
at $\epsilon_{\rm Q}$.)

\begin{figure}
\hspace*{0.5cm}
\psfig{file=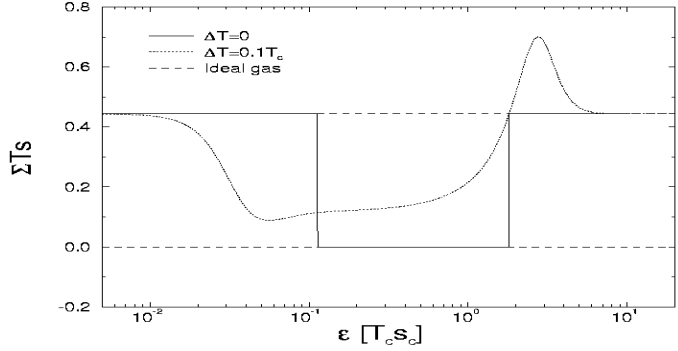,width=3.8in,height=2in,angle=-90}
\caption{The quantity $\Sigma$ (times $Ts$) as a function of
$\epsilon$ for $\Delta T=0$ (solid line), $\Delta T=0.1\, T_{\rm c}$
(dotted line), and the ideal hadron gas equation of state (dashed line).}
\label{ris:Fig9}
\end{figure}

Another quantity of interest is $\Sigma$, which determines whether
matter is thermodynamically normal or anomalous. Figure \ref{ris:Fig9}
shows this quantity (times $Ts$) as computed from (\ref{ris:Sigma})
for the three equations of state studied here.
For $\Delta T=0$, matter becomes anomalous in the mixed phase, the other
two equations of state are thermodynamically normal everywhere.
(Strictly speaking, $\Sigma=0$ only vanishes in the mixed phase, but does
not become negative. This is, however, sufficient for the formation
of stable rarefaction shock waves.)

\begin{figure}
\hspace*{2cm}
\psfig{file=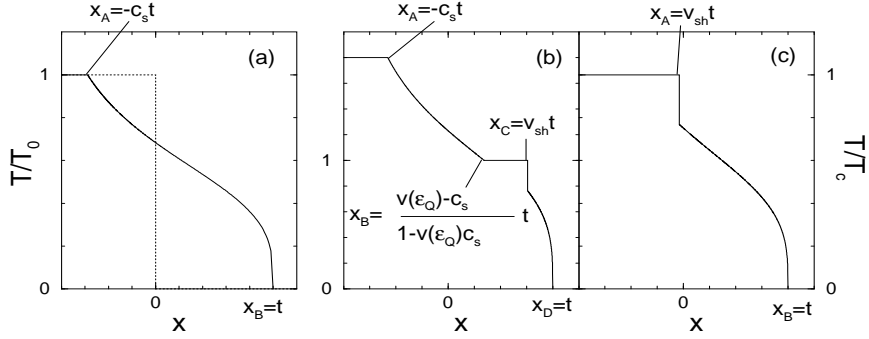,width=1.5in,height=3.5in,angle=-90}
\caption{Temperature profiles for the expansion of semi-infinite 
matter into vacuum. (a) Ideal hadron gas equation of state, the
dotted line indicates the initial state, the temperature is normalized
to the initial temperature $T_0$.
(b,c) Equation of state with $\Delta T=0$, in (b) the initial
energy density is well above $\epsilon_Q$, in (c) it is just below
$\epsilon_Q$. The temperature in (b,c) is normalized to the critical
temperature $T_{\rm c}$.}
\label{ris:Fig10}
\end{figure}

Let us now consider the one-dimensional expansion of semi-infinite matter
into the vacuum. Figure \ref{ris:Fig10} shows temperature profiles 
for (a) the expansion of an ideal gas and (b,c) for the expansion with
the $\Delta T=0$ equation of state. 
In (b) the initial energy density of semi-infinite matter is chosen 
to be well above $\epsilon_{\rm Q}$, the phase boundary between
the quark-gluon and the mixed phase, in (c) the initial energy 
density is just below $\epsilon_{\rm Q}$.
The dotted line in (a) indicates the
initial temperature profile for all cases.
The initial profile indicates a discontinuity at $x=0$ which separates
two regions of constant flow, the semi-infinite slab of matter at rest to
the left ($x\leq0$), and the vacuum to the right ($x>0$). This initial
condition is in fact a special case of the Riemann problem discussed in
Subsection 3.3.
From general arguments (see above) the solution at $t>0$ can
only be a simple wave, connecting these two regions of constant flow.
For the ideal hadron gas which is thermodynamically normal matter, 
we have seen above that this simple wave must be a continuous
rarefaction wave, in this case moving to the right. As mentioned above,
for such a wave the Riemann invariant ${\cal R}_+ = {\rm const.}$ everywhere, 
cf.\ (\ref{ris:Riemann}), from which we deduce the relationship between the 
fluid rapidity $y$ and the energy density $\epsilon$ on the rarefaction wave:
\begin{equation}
y(\epsilon) = - \frac{c_{\rm s} }{1+c_{\rm s}^2}\, 
\ln \frac{\epsilon}{\epsilon_0} \enspace ,
\end{equation}
where we have used the fact that for the ideal hadron gas equation of
state $p=c_{\rm s}^2 \epsilon$ and that the initial fluid rapidity of the
semi-infinite slab is zero, $y_0 = 0$. The fluid velocity on the 
rarefaction wave is then given by $v(\epsilon) = \tanh y(\epsilon)$.
Finally, the position at which one finds a given energy density $\epsilon$
at time $t$ can be deduced by integrating (\ref{ris:1charact}) for the
non-trivial ${\cal C}_-$ characteristics:
\begin{equation} \label{ris:xt}
x(\epsilon) = \frac{v(\epsilon)-c_{\rm s}}{1-v(\epsilon)\, c_{\rm s}} \, t
\enspace ,
\end{equation}
where we have used the fact that the initial position of the simple wave
is at $x=0$ and that $c_{\rm s}={\rm const.}$ for the ideal
hadron gas equation of state (we have assumed that the hadron gas consists
of massless, i.e., ultrarelativistic pions, for which $c_{\rm s}^2 = 1/3$).
Equation (\ref{ris:xt}) tells us that the rarefaction wave moves with
sound velocity into the semi-infinite slab of matter (to the left),
$x_{\rm A} = - c_{\rm s} t$, and with the velocity of light into the vacuum
(to the right), $x_{\rm B} = t$.

The expansion in the case of a first order phase transition, $\Delta T=0$,
proceeds similarly, with the exception that in the region of energy
densities corresponding to the mixed phase, matter is thermodynamically
anomalous, cf.\ Fig.\ \ref{ris:Fig9}, such that from Table \ref{ris:table1}
we conclude that the stable wave pattern is not 
a continuous rarefaction wave, but a rarefaction shock wave. Thus, as long
as matter is in the (thermodynamically normal) quark-gluon phase, the
expansion will proceed as a continuous rarefaction wave as 
in Fig.\ \ref{ris:Fig10} (a), but upon entering the mixed phase (energy
density $\epsilon_{\rm Q}$, temperature $T_{\rm c}$) a rarefaction shock
wave will form. The state of matter emerging
from this shock wave is determined from the shock equations
as described in the previous subsection, i.e., it corresponds to the
Chapman--Jouguet point on the Taub adiabat with center located at the
phase boundary between quark-gluon and mixed phase (for more details, see
\cite{ris:dhr}). Then, also the velocity of the shock $v_{\rm sh}$
in the calculational frame is determined. 
In general $v_{\rm sh}$ and the velocity of matter at the base
of the continuous rarefaction wave are not equal. This leads
to the formation of a plateau of constant flow between $x_{\rm B}$ and
$x_{\rm C}$.
The state of matter at the Chapman--Jouguet point corresponds 
to thermodynamically normal hadronic matter, so that the further 
expansion has to proceed as a continuous rarefaction wave. 
The emerging wave pattern is shown in Fig.\ \ref{ris:Fig10} (b).

The only difference between Fig.\ \ref{ris:Fig10} (b) and (c) is that
the initial energy density in (c) is already below $\epsilon_{\rm Q}$,
i.e., in the region corresponding to mixed phase. Therefore, the
expansion starts out with a rarefaction shock wave, from which matter
emerges at the Chapman--Jouguet point of the respective
Taub adiabat with center corresponding to the initial state of matter.
(Note that this Taub adiabat differs from the one in (b), since
their centers are different.) Further expansion proceeds as a continuous
rarefaction wave in hadronic matter.

This completes the discussion of the expansion of semi-infinite matter
into vacuum and prepares us to understand the Landau model which is
subject of the next subsection.

\subsection{The Landau Model}

\begin{figure}
\psfig{file=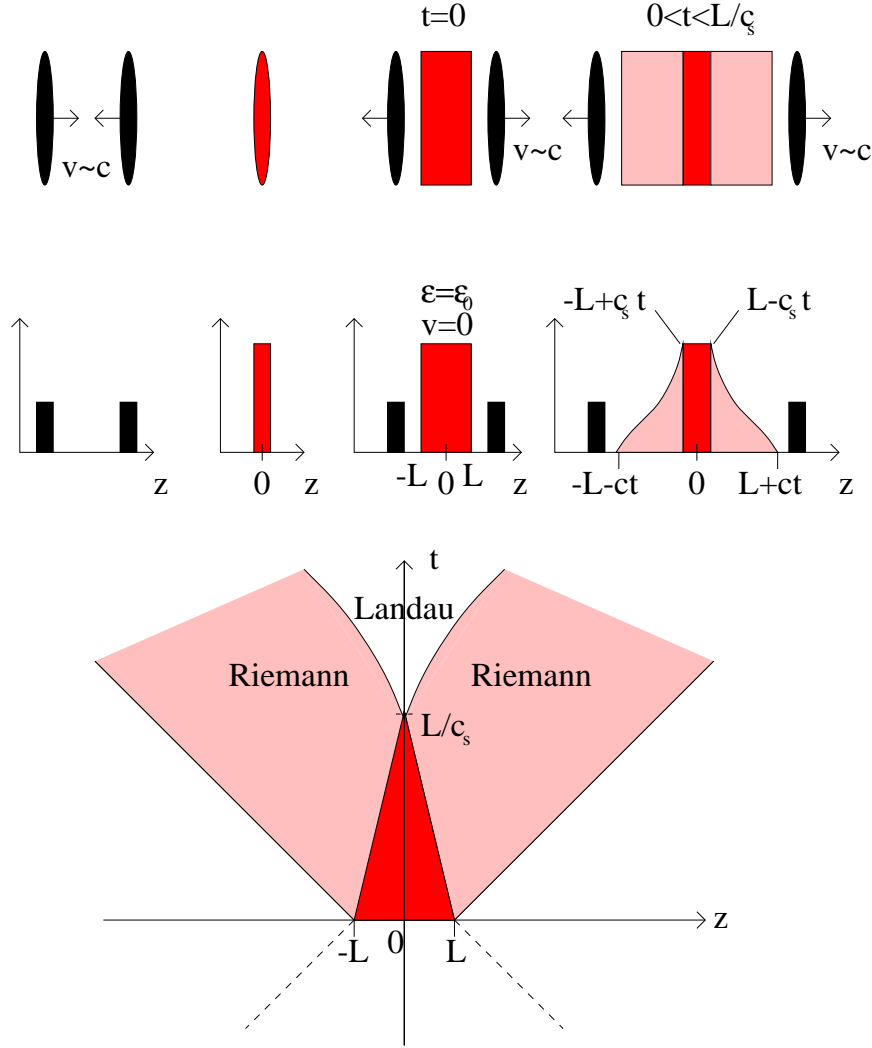}
\caption{The Landau model for nuclear collisions. See text for details.}
\label{ris:Fig11}
\end{figure}

The Landau model is historically the first case where fluid dynamics
was applied to describe -- at that time -- hadron-hadron 
collisions \cite{ris:landau}. Its main focus of application 
nowadays is, of course, nucleus-nucleus collisions.
The main ideas are summarized in Fig.\ \ref{ris:Fig11}.
Imagine two nuclei colliding at ultrarelativistic velocities
in their center of mass. The nuclei are Lorentz-contracted to a
``pancake-like'' shape. In the moment of impact, nuclear matter
becomes highly excited (the detailed microscopic processes which happen
during this stage are of no concern for the following).
In the limit that the velocities of the nuclei $v \rightarrow 1$,
there will be no baryon stopping (due to the limited stopping
power of nuclear matter), i.e., the baryon charges will pass through
each other unscathed, leaving highly excited, net baryon-free
matter in their wake. Due to Lorentz contraction, the initial extension 
$2\, L$ in $z$ direction of this slab of highly excited matter 
is much smaller than the transverse size of the 
slab, such that the expansion will proceed mainly in the longitudinal
direction and is thus essentially one-dimensional.
The Landau model assumes that the slab has no initial collective velocity and
that rapid thermalization takes place which is completed at $t=0$.
It is also assumed that the equation of state has the simple
ultrarelativistic form $p=c_{\rm s}^2 \epsilon$, $c_{\rm s}^2 = {\rm const.}$,
i.e., that matter is thermodynamically normal for all $\epsilon$.
(The original idea of Landau actually was that the baryons are
immediately stopped in the collision through 
compressional shock waves. Data from heavy-ion experiments at
BNL-AGS and CERN-SPS prove that this picture is unrealistic, due
to the aforementioned finite stopping power of nuclear matter.
However, since the collision is ultrarelativistic, the thermal
energy in the highly excited slab is much larger than the chemical
energy associated with the conservation of baryon charge. Therefore,
to good approximation, $\mu_{\rm B} = n_{\rm B} = 0$, and the further
evolution of the slab will be identical to what is discussed here.)

For $t>0$, the slab starts to expand. As in the expansion of semi-infinite
matter discussed in the previous subsection, rarefaction waves will
form. For thermodynamically normal matter, these are continuous
(Riemann) rarefaction waves which travel into the slab with sound velocity.
Therefore, they will meet
at the center of the slab (here chosen to be the origin $z=0$) at a 
time $L/c_{\rm s}$. For times $t> L/c_{\rm s}$, these waves overlap and
the solution becomes more complicated. In a region near the light cone,
the solution will remain a Riemann rarefaction wave, therefore we term
this region the {\em Riemann region}. 
In the center where the Riemann rarefaction waves overlap, however,
the solution is no longer a simple wave (indeed, only two regions
of constant flow {\em have\/} to be connected by a simple wave 
\cite{ris:courant}, for two simple waves no such theorem exists).
For $c_{\rm s}^2 = {\rm const.}$ the solution can still be given
in closed analytic form \cite{ris:landau}, although the derivation is
rather complicated. However, since two of our equations of state do not have
constant velocities of sound, we have to resort to numerical
solution methods, such as the relativistic HLLE discussed above. 
In principle, numerical algorithms can deal with arbitrary (physically
reasonable) equations of state, and are therefore well able to handle
this problem (although one should test them thoroughly for test cases
where analytical solutions are known \cite{ris:dhr,ris:dhr2}).

\begin{figure}
\psfig{file=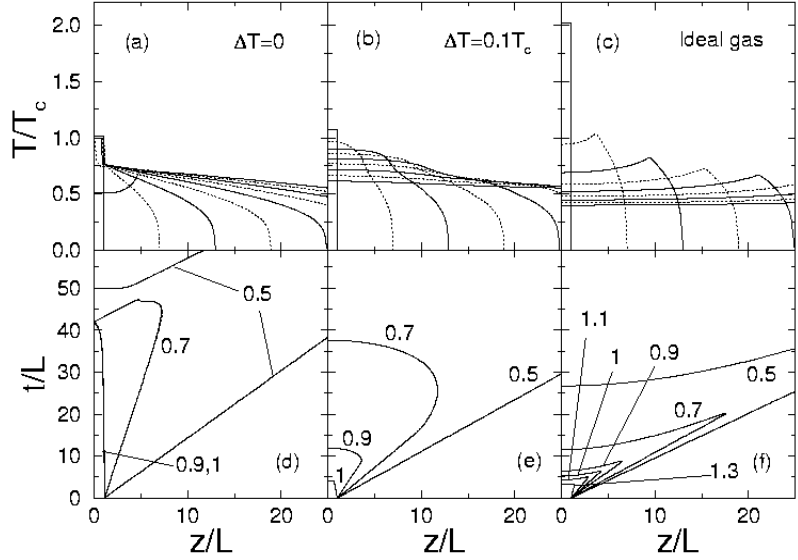,width=4.5in,height=3.5in,angle=-90}
\caption{Expansion in the Landau model for $\Delta T=0$ (a,d), $\Delta T
= 0.1\, T_{\rm c}$ (b,e), and the ideal gas equation of state (c,f).
(a--c) show temperature profiles for different times, (d--f) show
the corresponding isotherms in the $t-z$ plane (numbers are temperatures
in units of $T_{\rm c}$). The initial energy density is
$\epsilon_0 = 1.875\, T_{\rm c}s_{\rm c}$ in all cases.}
\label{ris:Fig12}
\end{figure}

In Fig.\ \ref{ris:Fig12} numerical solutions for the Landau model are
presented for the three different equations of state of Fig.\ \ref{ris:Fig8}.
The initial energy density is $\epsilon_0 = 1.875\, T_{\rm c} s_{\rm c}$
which is slightly larger than $\epsilon_{\rm Q}$.
In Figs.\ \ref{ris:Fig12} (a--c) temperature profiles are shown for
different times $t$ and for the $z>0$ half plane (the solution in
the other half plane is the respective mirror image).
For $\Delta T=0$, Fig.\ \ref{ris:Fig12} (a), one clearly observes the
rarefaction shock wave which, for this initial energy density is
almost stationary. Hadronic matter is expelled from the shock until the
energy in the interior of the slab decreases below 
$\epsilon_{\rm H}$ and the shock vanishes.
For $\Delta T=0.1 \, T_{\rm c}$, Fig.\ \ref{ris:Fig12} (b), 
no shock is formed, although the variation of the velocity of sound in the 
mixed phase, Fig.\ \ref{ris:Fig8} (d), leads to shapes for the continuous 
rarefaction waves which differ strongly from those for a constant velocity 
of sound, Fig.\ \ref{ris:Fig12} (c). Note the kink in the temperature
profiles in the latter case which indicate the position where the
Landau solution matches to the Riemann rarefaction wave.
Note also the difference in the initial temperatures for the three cases
although the initial energy density is the same. 
This is a consequence of the different number
of degrees of freedom for the three equations of state
at high energy densities.

In Figs.\ \ref{ris:Fig12} (d--f) corresponding isotherms are shown in 
the $t-z$ plane. The most pronounced feature is that due
to the small propagation 
velocity of the rarefaction wave, the system stays hot for a much longer 
time span for the $\Delta T=0$ equation of state, Fig.\ \ref{ris:Fig12} (d), 
than for the ideal gas, Fig.\ \ref{ris:Fig12} (f). 
This is in agreement with the general argument presented earlier that
the softening of the equation of state in the mixed phase region leads
to a reduced expansion tendency and thus to a ``stalled'' expansion
of the system. The softening of the equation of state
is also the reason why the expansion for the $\Delta T=0.1\, T_{\rm c}$
equation of state, Fig.\ \ref{ris:Fig12} (e), 
is delayed in comparison to the ideal gas case, although
no rarefaction waves are formed. For a quantitative analysis of the
delayed expansion in the Landau model see \cite{ris:dhrmg}.

\subsection{The Bjorken Model}

One of the main assumptions of Landau's model is that the initial
collective velocity of the slab of excited matter vanishes. 
However, this cannot be quite true on account of the following argument.
In the limit $v\rightarrow 1$, the size of the nuclei in longitudinal
direction goes to zero, and there is no scale in the problem at all.
In this case, the collective velocity of matter in the slab {\em has\/} 
to be of the scaling form $v=z/t$. The consequences of this special form 
for the longitudinal fluid velocity were first investigated in 
\cite{ris:cooper,ris:chiu}, again with respect to possible applications
in hadron-hadron collisions. Bjorken \cite{ris:bjorken}
was the first to discuss it in the framework of nuclear collisions.

\begin{figure}
\hspace*{1cm}
\psfig{file=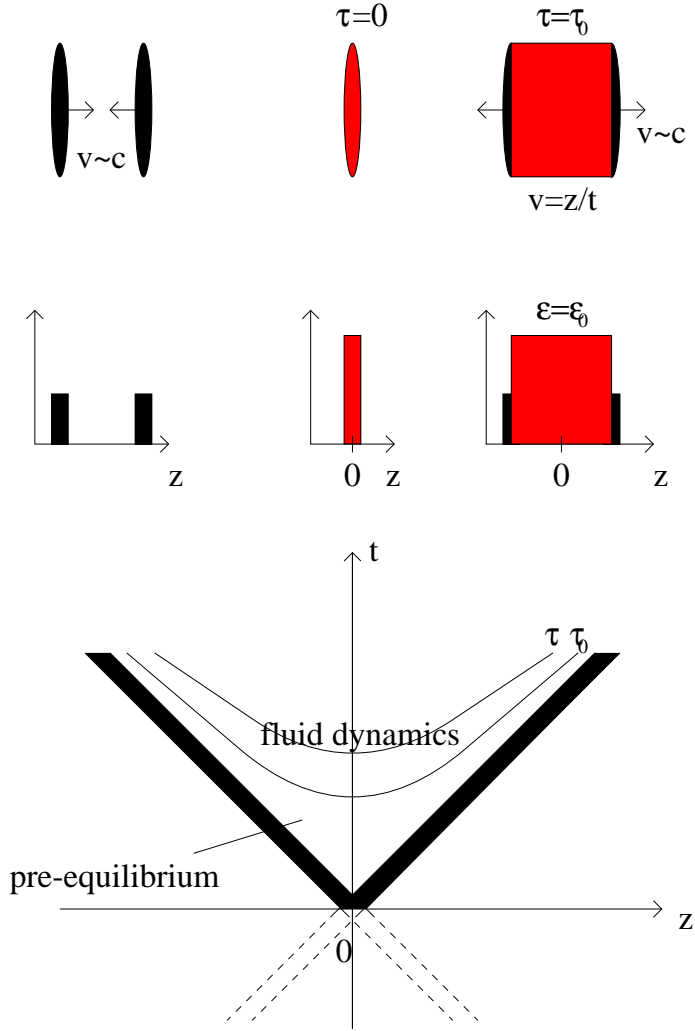}
\caption{The Bjorken model for nuclear collisions. See text for details.}
\label{ris:Fig13}
\end{figure}

The main ideas of the so-called {\em Bjorken model\/} are summarized
in Fig.\ \ref{ris:Fig13}. As in the Landau model, two ultrarelativistic,
Lorentz-contracted nuclei collide at $z=0$ and $t=0$ (the moment of
complete overlap) in the center of mass frame of the collision.
Due to the limited amount of nuclear stopping power, the baryon
charges keep on moving along the light cone, while microscopic collision
processes (the nature of which is of no concern for the following) lead to
the formation of a region of highly excited, net charge free matter in the wake
of the nuclei. In contrast to the Landau model, however, the collective
velocity in this region is of the scaling form $v=z/t$. The region
of highly excited matter is supposed to rapidly equilibrate locally within a
time span $\tau_0$ (which is of the order of a fm or less), and the further
evolution of the system can be described in terms of ideal fluid dynamics.
One important point is that, due to the absence of a scale, physics has to
be the same for matter at different longitudinal coordinate $z$ if
compared at the same {\em proper time\/} $\tau = t \sqrt{1-v^2} =
\sqrt{t^2-z^2}$. (Such curves of constant $\tau$ describe 
hyperbola in space-time.) Thus, the initial thermodynamic state of all 
fluid elements is the same at the same {\em proper time\/} $\tau_0$. 

If the longitudinal velocity profile is enforced by the scaling argument,
the fluid-dynamical solution simplifies in fact considerably. To see
this, change the variables $t,\,z$ in 
the conservation laws for one-dimensional longitudinal
motion in the absence of conserved charges,
\begin{equation} \label{ris:bjmotion}
\partial_t \, T^{00} + \partial_z\, T^{z0} = 0 \,\,\, ,\,\,\,\,
\partial_t \, T^{0z} + \partial_z\, T^{zz} = 0 \enspace ,
\end{equation}
to the variables $\tau = \sqrt{t^2 - z^2}$, which is the proper time
of a fluid element, and $\eta = {\rm Artanh}\, v = {\rm Artanh}[ z/t]$,
which is the rapidity of a fluid element. Then, the coupled system of partial
differential equations (\ref{ris:bjmotion}) decouples:
\begin{eqnarray} \label{ris:dedtau}
\left. \frac{\partial \epsilon}{\partial \tau} 
\right|_\eta + \frac{\epsilon +p}{\tau} & = & 0 \enspace , \\
\left. \frac{\partial p}{\partial \eta} \right|_\tau & = & 0 \enspace .
\label{ris:dpdeta}
\end{eqnarray}
The second equation (\ref{ris:dpdeta}) has the interesting consequence
that there is no pressure gradient between adjacent fluid elements
(the one at $\eta$ and the one at $\eta + \D \eta$). At first glance
this would seem to indicate that there is no expansion of the fluid at all.
This, however, is not true, since the fluid velocity is certainly finite,
$v=z/t$. The answer is that the new coordinates $(\tau,\eta)$ 
already take the scaling expansion into account: 
a fluid element at $\eta$ with a width $\Delta \eta$ 
in fact ``grows'' in longitudinal direction by an
amount $\D z = \D t\, \Delta \eta$ during the time span $\D t$ . 

Another consequence of (\ref{ris:dpdeta}) is derived from the
Gibbs--Duhem relation:
\begin{equation} \label{ris:GD}
\left.  \frac{\partial p}{\partial \eta} \right|_\tau  = 
s \left.  \frac{\partial T}{\partial \eta} \right|_\tau + 
\sum_{i=1}^n n_i 
\left.  \frac{\partial \mu_i}{\partial \eta} \right|_\tau =0 \enspace .
\end{equation}
This equation means that for vanishing conserved 
charges $n_i = 0$, $i=1, \ldots,n$, the temperature has to be constant 
along curves of constant $\tau$, i.e., along the space-time hyperbola shown
in Fig.\ \ref{ris:Fig13} ($\eta$ varies along these curves). 
In the general case of non-zero net charges, 
however, only the particular combination of charge densities,
entropy density, and derivatives of $T$ and the $\mu_i$ 
appearing in (\ref{ris:GD}) has to vanish along curves of constant $\tau$.
Equation (\ref{ris:dpdeta}) represents the principle of ``boost invariance''
commonly associated with the Bjorken model: at constant $\tau$
the pressure is independent of the longitudinal rapidity, i.e., 
it is the same in fluid elements with different $\eta$, or in other
words, it does not change if one performs a longitudinal boost to a different 
reference frame. This is a consequence of the scaling form for 
the longitudinal velocity.

Equation (\ref{ris:dedtau}) also has an interesting consequence.
With the first law of thermodynamics, one derives as usual the
conservation of the entropy current which now takes the form
\begin{equation}
\left.  \frac{\partial s}{\partial \tau} \right|_\eta +\frac{s}{\tau} = 0
\enspace ,
\end{equation}
which can be immediately integrated to give
\begin{equation} \label{ris:stau}
s\, \tau = s_0 \tau_0 = {\rm const.} 
\end{equation}
at constant $\eta$. The constant may in principle
differ for different $\eta$, but since the initial thermodynamic state
along $\tau_0$ was the same for all $\eta$, that constant will 
also be the same for all $\eta$ at other $\tau > \tau_0$.
Equation (\ref{ris:stau}) is interesting because it tells us
that the entropy density decreases inversely proportional to
$\tau$ {\em independent of the equation of state\/} of the fluid.
The time evolution for energy density, pressure, or temperature might
depend on the equation of state, but not the one for the entropy density.

\begin{figure}
\vspace*{0.5cm}
\psfig{file=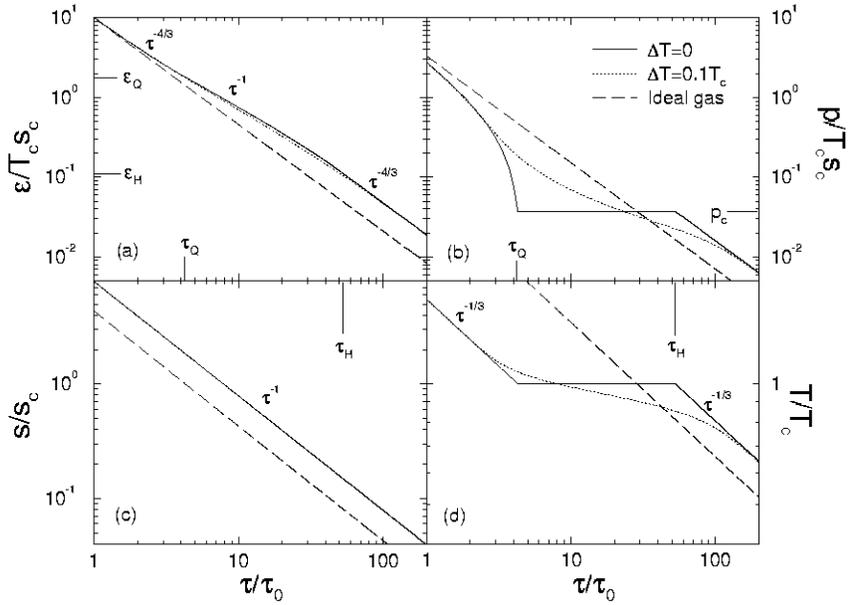,width=4.5in,height=3.5in,angle=-90}
\caption{Proper time evolution for (a) energy density, (b) entropy density,
(c) pressure, and (d) temperature in the Bjorken model for nuclear 
collisions (longitudinal expansion only). Solid line: $\Delta T=0$,
dotted line: $\Delta T = 0.1\, T_{\rm c}$, dashed line: ideal gas equation
of state. The initial energy density is $\epsilon_0 = 10\, T_{\rm c}
s_{\rm c}$.}
\label{ris:Fig14}
\end{figure}

This is confirmed in Fig.\ \ref{ris:Fig14}, where the evolution of
(a) the energy density, (b) the entropy density, (c) the pressure, and
(d) the temperature is shown as a function of proper time $\tau$ for the
three equations of state ($\Delta T=0$, $\Delta T=0.1\, T_{\rm c}$, and
the ideal hadron gas). Note that in the quark-gluon as well as the
hadron phase, where $p \sim c_{\rm s}^2 \, \epsilon$ with $c_{\rm s}^2 =1 /3$, 
Eq.\ (\ref{ris:dedtau}) yields
\begin{equation}
\epsilon \sim \tau^{-4/3} \enspace ,
\end{equation}
For the $\Delta T=0$ equation of state, $p=p_{\rm c} = {\rm const.}$ in
the mixed phase, and (\ref{ris:dedtau}) yields the cooling law
\begin{equation}
\epsilon \sim \tau^{-1} \enspace .
\end{equation}
This is interpreted as follows. The longitudinal scaling expansion
dilutes the system $\sim \tau^{-1}$. If no mechanical work is performed,
like in the mixed phase where $\D p \equiv 0$,
only this geometrical dilution determines the (proper) time evolution of
the energy density. In the phase where $\D p = c_{\rm s}^2 \, \D \epsilon$,
however, additional mechanical work is performed, and the system cools 
faster, $\epsilon \sim \tau^{-(1+c_{\rm s}^2)} = \tau^{-4/3}$.
The faster cooling 
is confirmed studying the temperature evolution, Fig.\ \ref{ris:Fig14}
(d). For $p=c_{\rm s}^2\, \epsilon$, $c_{\rm s}^2= {\rm const.}$, and
vanishing net charges, one deduces from 
$\D p = c_{\rm s}^2\, \D \epsilon = s\, \D T = (\epsilon + p)\, \D T/ T =
(1+c_{\rm s}^2)\, \epsilon \,\D T/T$, that $\epsilon \sim T^{1+c_{\rm s}^{-2}}$
and consequently, in the hadron and quark-gluon phase
\begin{equation}
T \sim \tau^{-1/3} \enspace,
\end{equation}
while in the mixed phase one deduces from $\D p = s \,\D T \equiv 0$ that
\begin{equation}
T = {\rm const.} \enspace. 
\end{equation}
This expectation is confirmed in Fig.\ \ref{ris:Fig14} (d).

Of course, in reality the expansion of the system will not only be
purely longitudinal. The ``Bjorken cylinder'' will also expand transversally.
The principle of boost invariance allows us to focus on the transverse
expansion at $z=\eta=0$ only, and reconstruct the fluid properties at a 
different $\eta$ by performing a longitudinal boost with boost rapidity
$\eta$.
For the sake of simplicity, let us assume that the system is cylindrically
symmetric in the transverse direction and that the initial energy density
profile is of the form
\begin{equation} \label{ris:bjcylin}
\epsilon(\vec{r},\tau_0, \eta=0) = \epsilon_0\, \Theta(R-|\vec{r}|)
\enspace ,
\end{equation}
where $R$ is the transverse radius of the Bjorken cylinder. In
cylindrical coordinates and at $z=\eta=0$, the conservation equations
read ($T^{00} \equiv E,\, T^{0r} \equiv M,\, v_r \equiv v$):
\begin{eqnarray} \label{ris:bjcylE}
\partial_t\, E + \partial_r \, \left[ (E+p)v \right] & = & - 
\left( \frac{v}{r} + \frac{1}{t} \right) (E + p) \enspace , \\
\partial_t \, M + \partial_r\, (Mv + p) & = & -\left( \frac{v}{r}
+ \frac{1}{t} \right) M \enspace .\label{ris:bjcylM}
\end{eqnarray}
Although these equations have no longer a simple analytical solution, 
the assumption of cylindrical symmetry has reduced the originally 
three-di\-men\-sio\-nal problem to an effectively one-dimensional problem.
Indeed, for vanishing right-hand sides the solution of
(\ref{ris:bjcylE}), (\ref{ris:bjcylM}) with the initial condition
(\ref{ris:bjcylin}) is identical to the one of the Landau model with
the substitutions $z\rightarrow r$ and $L \rightarrow R$. The right-hand
sides just lead to an additional reduction of $E$ and $M$ 
from the cylindrical geometry, $v/r$, and from longitudinal scaling,
$1/t$.

\begin{figure}
\psfig{file=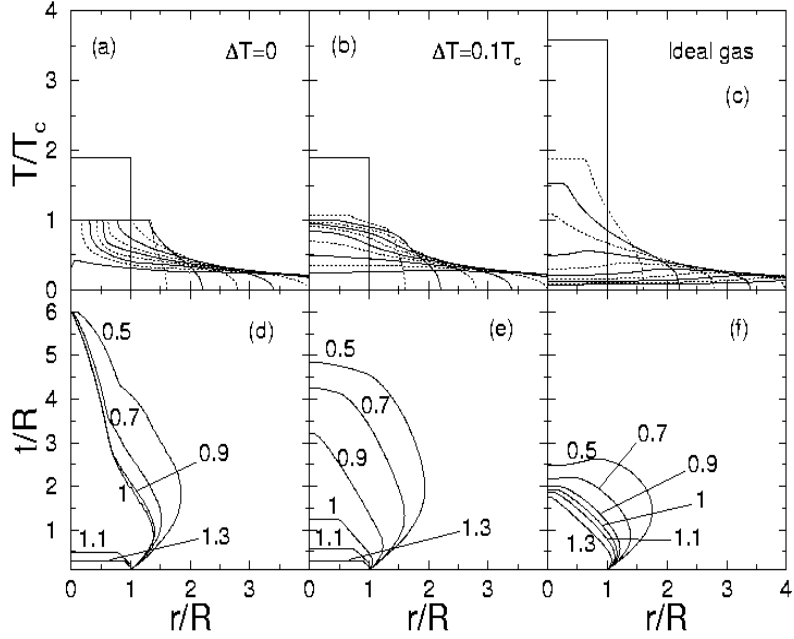,width=4.5in,height=3.8in,angle=-90}
\caption{Transverse expansion of the Bjorken cylinder
for $\Delta T=0$ (a,d), $\Delta T
= 0.1\, T_{\rm c}$ (b,e), and the ideal gas equation of state (c,f).
(a--c) show temperature profiles for different times, (d--f) show
the corresponding isotherms in the $t-r$ plane (numbers are temperatures
in units of $T_{\rm c}$). The initial energy density is
$\epsilon_0 = 18.75\, T_{\rm c}s_{\rm c}$ in all cases.}
\label{ris:Fig15}
\end{figure}

This observation, combined with the method of operator splitting 
discussed previously, suggests the following simple solution scheme
(also known as Sod's method \cite{ris:holt,ris:sod}):
equations (\ref{ris:bjcylE}), (\ref{ris:bjcylM}) are of the type
\begin{equation}
\partial_t \, U + \partial_x \, F(U) = - G(U) \enspace.
\end{equation}
The operator splitting method allows to construct the solution by first
solving the one-dimensional {\em partial\/} differential equation
\begin{equation}
\partial_t \, U + \partial_x \, F(U) = 0 \enspace,
\end{equation}
(for instance with the relativistic HLLE scheme discussed above),
which yields a {\em prediction\/} $\tilde{U}$ for the true solution $U$.
In a second step one {\em corrects\/} this prediction by solving
the {\em ordinary\/} differential equation
\begin{equation}
\frac{\D U}{\D t} = - G(U) \enspace ,
\end{equation}
which is numerically realized as \cite{ris:dhrmg2}
\begin{equation}
U = \tilde{U} - \Delta t \, G(\tilde{U}) \enspace .
\end{equation}
The transverse expansion of the Bjorken cylinder at $z=0$ is shown in
Fig.\ \ref{ris:Fig15} for $\tau_0=0.1\, R$ and $\epsilon_0 = 18.75\,
T_{\rm c}s_{\rm c}$. One immediately recognizes the qualitative similarities
with the Landau expansion, like the delay in the expansion for the two
equations of state with a (phase) transition as compared to the expansion
with an ideal hadron gas equation of state.
The additional geometrical dilution, however,
leads in general to a faster cooling overall and 
quantitatively different shapes
for the temperature profiles and the isotherms in the $t-r$ plane.

\begin{figure}
\vspace*{0.5cm}
\hspace*{0.5cm}
\psfig{file=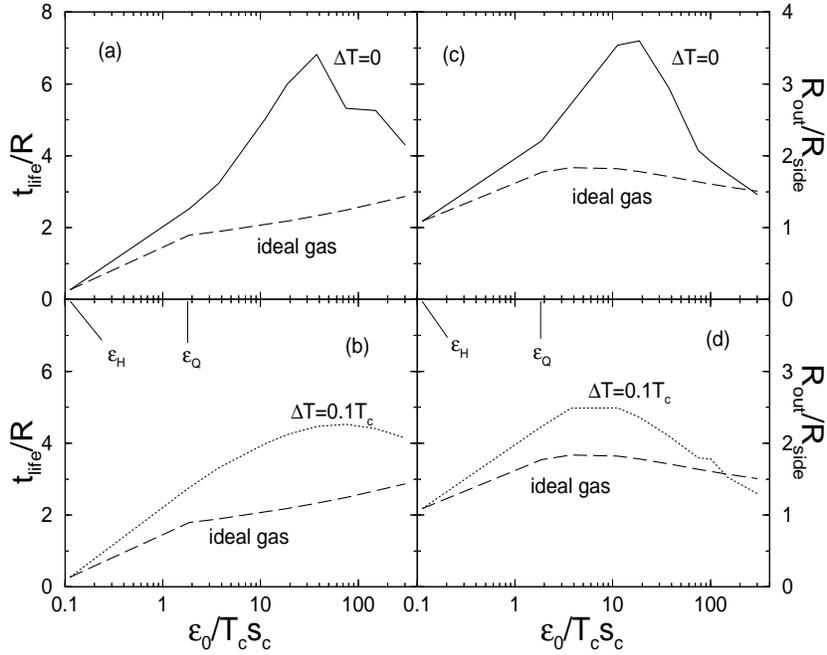,width=2.7in,height=3.8in,angle=-90}
\caption{Lifetime of the system as a function of $\epsilon_0$ for
the Bjorken cylinder expansion, $\tau_0 = 0.1\, T_{\rm c}$.
(a) $\Delta T =0$ (solid) vs.\ ideal hadron gas (dashed),
(b) $\Delta T =0.1\, T_{\rm c}$ (dotted) vs.\ ideal hadron gas (dashed). 
(c,d) the corresponding ratio $R_{\rm out}/R_{\rm side}$.}
\label{ris:Fig16}
\end{figure}

Let us further quantify the time delay in the expansion induced by the 
transition in the equation of state. In general, the system will decouple
into free-streaming particles once the temperature drops below a certain
``freeze-out'' temperature $T_{\rm fo}$, see Section 5 below. From
comparison with experimental data, this freeze-out
temperature is estimated to be on the order of 100 MeV.
Let us therefore define a ``lifetime'' of the system as the time 
when the $T=0.7\, T_c$ isotherm crosses the origin at $r=0$ in Figs.\
\ref{ris:Fig15} (d--f). This lifetime is shown in Figs.\ \ref{ris:Fig16} 
(a,b) as function of the initial energy density $\epsilon_0$ of the 
cylinder. One observes a maximum of the such defined lifetime at
initial energy densities around $40\, T_{\rm c}s_{\rm c} \sim 30\,
{\rm GeVfm}^{-3}$. At these initial energy densities, 
the prolongation of the lifetime over the
respective ideal hadron gas value is about a factor
of 2 (for $\Delta T=0.1\, T_{\rm c}$) to 3 (for $\Delta T=0$).

The prolongation of the lifetime is due to the softening of the equation
of state in the phase transition region. 
It is, however, interesting that the maximum in the lifetime does not occur
around initial energy densities corresponding to $\epsilon_{\rm Q}$ 
(as is the case in the Landau model \cite{ris:dhrmg}), but at much 
larger initial energy densities.
The reason for this is the strong longitudinal dilution of
the system on account of the scaling profile $v_z = z/t$. In order to
see a large effect of the softening of the equation of state in the
phase transition region on the expansion dynamics, the transverse (Landau-like)
expansion has to be the dominant cooling mechanism for the system. 
The Bjorken scaling expansion does not account for the reduced expansion
tendency of the system in the transition region, it {\em enforces\/} an
expansion velocity $v_z=z/t$ irrespective of the equation of state.
In order to have the transverse expansion dominate the cooling of the
system, one has to start the expansion
at higher initial energy densities such that
the system spends enough time in the mixed phase for the (slow) rarefaction
shock to reach the origin. The initial energy density in Fig.\ 
\ref{ris:Fig15} was intentionally selected to maximize this effect.

Initial energy densities on the order of $10-30\, {\rm GeVfm}^{-3}$
are expected to be reached at the RHIC collider. In order to experimentally
observe the prolongation of the lifetime as seen in Figs.\ \ref{ris:Fig16},
one has to find a corresponding experimental observable. An obvious
candidate is the ratio of the ``out'' to the ``side'' radius of two-particle
correlation functions. The ``out'' radius is proportional to the
duration of particle emission from a source, while the ``side'' radius
is proportional to the transverse dimension of the source (cf.\
\cite{ris:uli} for a very detailed, pedagogical discussion). 
Since the transverse radius of the source is approximately the same in
all cases, cf.\ Fig.\ \ref{ris:Fig15} (a--c), the ratio 
$R_{\rm out}/R_{\rm side}$ seems to be a good generic 
measure for the lifetime. Moreover, in forming the ratio the dependence
on the overall (unknown) spatial size of the source as well as 
effects from the collective expansion are expected to cancel.
The ratio $R_{\rm out}/R_{\rm side}$ is plotted in Figs.
\ref{ris:Fig16} (c,d) for pions with mean transverse momenta 
$K_\perp=300$ MeV. (Details on how to compute this quantity can be found
in \cite{ris:dhrmg2,ris:dhrber}.) As one observes, $R_{\rm out}/R_{\rm side}$
nicely reflects the excitation function of the lifetime of the system.

\section{Freeze-Out}

\begin{figure}
\psfig{file=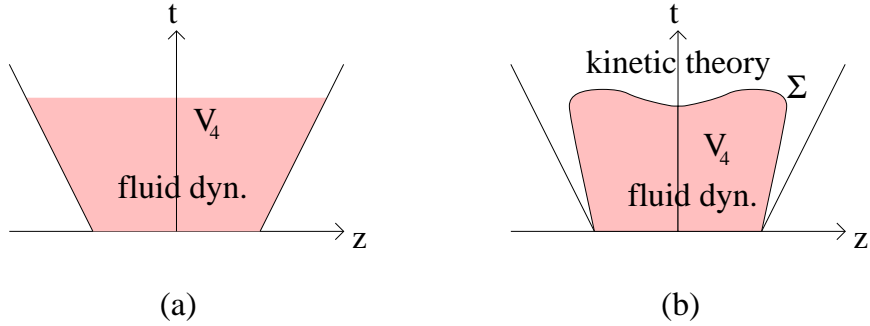}
\caption{(a) Conventional fluid-dynamical description in the whole
forward lightcone. (b) Fluid dynamics describes the evolution of the
system inside $V_4$, while kinetic theory describes the motion of the
frozen-out particles outside $V_4$.}
\label{ris:Fig17}
\end{figure}

In this section I discuss an up to date unsolved problem in the
application of relativistic fluid dynamics to describe nuclear collisions,
namely the so-called ``freeze-out'' process.
Given an initial condition, 
fluid dynamics describes the evolution of the system in the whole forward
lightcone, Fig.\ \ref{ris:Fig17} (a). However, as we have seen above,
at all times near the boundary to the vacuum, as well as everywhere
in the late stage of the evolution, the energy density becomes arbitrarily 
small, i.e., the system is rather cold and dilute. In this space-time region
the assumption of local thermodynamical equilibrium is no longer justified,
because the particle scattering cross section $\sigma$ is finite,
such that for small particle densities $n$ the particle scattering rate, 
$\Gamma \sim \sigma n$, becomes on the order of the inverse
system size, $\Gamma \sim R^{-1}$. At this point, the scattering rate
is too small to maintain local thermodynamical equilibrium and the particles
decouple from the fluid evolution. 
In this space-time region, a kinetic description for the
particle motion would be more appropriate. One should therefore not
solve fluid-dynamical equations in the whole forward lightcone, but only
inside a space-time region of sufficiently large energy and particle
densities, while outside this region, the particle motion should be described
by kinetic theory, Fig.\ \ref{ris:Fig17} (b).

The boundary $\Sigma$ between the two regions is determined by a 
criterion which compares local scattering rates with the system size,
as discussed above.
The obvious difficulty with this more realistic description of the
system's evolution is that this boundary 
has to be determined {\em dynamically}, i.e., not only has
one to allow for particles decoupling from the fluid, but also for the
reverse process of particles entering the fluid from the kinetic region.
(This can happen since the particles still, albeit rarely, collide 
in the kinetic region.) A consistent treatment of this problem is
rather complicated, since one has to solve kinetic in addition
to the fluid-dynamical equations. No serious attempt has been made so far.

Instead, the following approximate solution has been extensively employed:
\begin{enumerate}
\item One assumes that fluid dynamics gives a reasonable description for the
evolution of the system in the {\em whole\/} forward lightcone.
\item One determines the ``decoupling'' surface $\Sigma$ {\em a posteriori\/},
once the evolution of the fluid is known. 
\item The ``thickness'' of $\Sigma$ is assumed to be infinitesimal.
\item One assumes that particles crossing $\Sigma$ have {\em completely\/}
decoupled from the system, they stream freely towards the detectors without
any further collisional interaction (``freeze-out''). This means that
they do not change their momentum and energy once they have crossed
$\Sigma$.
\end{enumerate}
A very popular argument in order to determine $\Sigma$ is the following.
Since $n \sim T^3$, the scattering rate $\Gamma \sim T^3$ (for
constant cross section $\sigma$), i.e., if the temperature falls
below a certain so-called ``freeze-out'' temperature $T_{\rm fo}$, the
criterion $\Gamma \la R$ is fulfilled, and particles 
decouple from the system. In this case, $\Sigma$ is just given
by the isotherm $T=T_{\rm fo}$ (use of this argument was already 
made above in the discussion of the ``lifetime'' of the system).

Note that assumption 3.\ is a strong idealization and actually rather 
questionable, because in reality $\Sigma$ is a space-time region of finite 
thickness, inside which non-equilibrium, dissipative effects
become gradually more and more important (the more dilute the fluid
becomes), until ultimately all interactions between
particles cease and, when leaving $\Sigma$, they stream freely towards 
the detectors. 

Nevertheless, with the above assumptions, one can readily compute the 
single inclusive spectra of particles reaching the detector. Immediately 
before the particles decouple from the fluid evolution, i.e., before they 
cross $\Sigma$,
they are still in local thermodynamical equilibrium such that their phase space
distribution is given by $f_0(k,x)$, Eq.\ (\ref{ris:feq}). It is 
reasonable to assume that this phase space distribution is not changed
much when they move a small distance along their worldlines, which carries
them across $\Sigma$ into the region of free-streaming. 
In that region, however, there are no collisions which could further change
$f_0$. Therefore, the phase space distribution of ``frozen-out'' particles 
is (approximately) the same as in local equilibrium.
The total number of particles crossing a small surface element
$\D \Sigma$ of $\Sigma$ is then given by
\begin{equation}
N_\Sigma \equiv \D \Sigma_\mu N^\mu = \int \frac{\D^3 \vec{k}}{E}\, 
\D \Sigma \cdot k
\, f_0(k,x) \enspace ,
\end{equation}
$N^\mu$ being the (kinetic) particle number 4-current.
The {\em invariant momentum spectrum\/} of particles crossing that
surface element is consequently
\begin{equation}
E\, \frac{\D N_\Sigma}{\D^3 \vec{k}} = \D \Sigma \cdot k \, f_0(k,x)
\enspace .
\end{equation}
Finally, the invariant momentum spectrum (the {\em single inclusive\/}
spectrum) of particles crossing the {\em complete\/}
``freeze-out'' surface $\Sigma$ is
\begin{equation}
E \frac{\D N}{\D^3 \vec{k}} = \int_\Sigma \, 
E\, \frac{\D N_\Sigma}{\D^3 \vec{k}} = \int_\Sigma
\D \Sigma \cdot k \, f_0(k,x) \enspace .
\end{equation}
This equation is known as the {\em Cooper--Frye formula\/} \cite{ris:cooper},
and is used in almost all fluid-dynamical applications to heavy-ion collisions
to compute the single inclusive spectra of particles. 

\begin{figure}
\hspace*{0.5cm}
\psfig{file=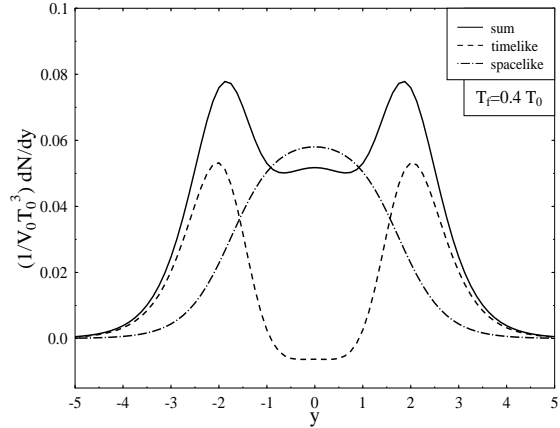,width=2.7in,height=3.8in,angle=-90}
\caption{The rapidity distribution for freeze-out along the $T_{\rm fo}
= 0.4\, T_0$ isotherm in the Landau model. 
Solid: full distribution, dotted: particles
from time-like parts of the isotherm, dash-dotted: particles 
from space-like parts of the isotherm.}
\label{ris:Fig18}
\end{figure}

There is, however, a problem with this formula \cite{ris:kyrill}.
For {\em time-like\/} surfaces, i.e., where the normal vector
$\D \Sigma_\mu$ is {\em space-like\/}, $\D \Sigma \cdot k$ may either
be positive or negative, depending on the value and direction of
$k^\mu$. In other words, the number of particles ``freezing out'' 
from a certain time-like surface element $\D \Sigma$ can become negative. 
This is clearly unphysical, since the number of particles decoupling
from the system must be positive definite. 
For {\em space-like\/} surfaces (with a {\em time-like\/} normal vector) as
well as for time-like surface elements where $\D \Sigma \cdot k >0$,
the Cooper-Frye formula gives a physically reasonable, positive definite 
result for the number of frozen-out particles.
This is illustrated in Fig.\ \ref{ris:Fig18} which shows the
rapidity distribution of particles (i.e., the invariant momentum spectrum
integrated over all transverse momenta) for massless particles decoupling
from a freeze-out isotherm $T_{\rm fo} = 0.4\, T_0$ in the Landau model
with a $p = \epsilon/3$ equation of state. One clearly notices the negative
particle numbers at midrapidity coming from the time-like parts of
the isotherm.

\begin{figure}
\hspace*{2cm}
\psfig{file=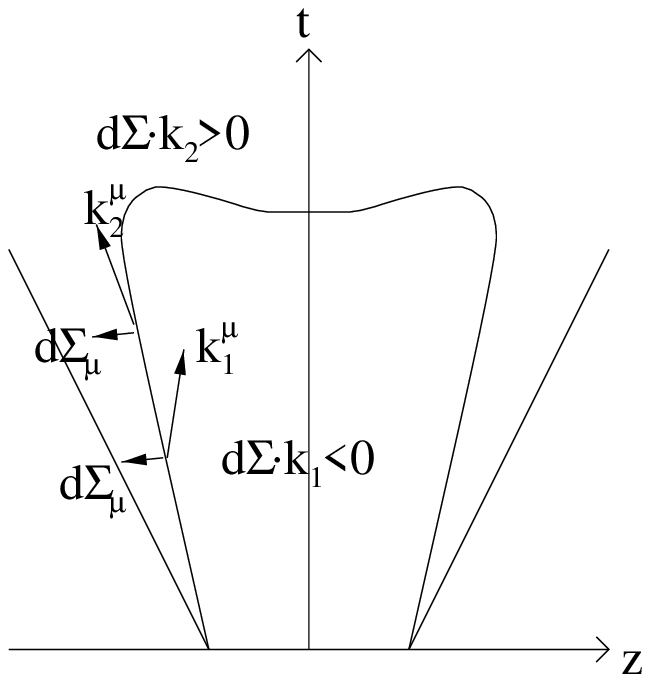}
\caption{Explanation for the negative number of frozen-out particles
in the Cooper-Frye formula.}
\label{ris:Fig19}
\end{figure}

This contradiction is readily resolved noting that the Cooper-Frye formula
does not really determine the number of particles decoupling from the
system, but merely the number of particle {\em worldlines crossing a
surface element\/} $\D \Sigma$ (and then integrated over the whole surface
$\Sigma$). For time-like surface elements, there is of course the possibility
that for certain $k^\mu$ the respective worldlines cross 
$\D \Sigma$ in the ``wrong'' direction, i.e., the momenta of the 
particles point back into the region of fluid, cf.\
Fig.\ \ref{ris:Fig19}. In particular, for the
$T_{\rm fo} = 0.4\, T_0$ isotherm, which moves {\em away\/} from the 
$t$-axis in the $t-z$ plane, those are particles with vanishing momentum
component in $z$ direction, because their worldlines are parallel to the 
$t$-axis.
Particles with $p^z=0$, however, also have vanishing longitudinal rapidity
$y=0$, and that is the reason why these negative particle numbers appear
at midrapidity in Fig.\ \ref{ris:Fig18}.
While this explains the negative contributions in the Cooper-Frye
formula, it also invalidates this formula as the correct prescription to
calculate the spectra of frozen-out particles, if parts of the decoupling
surface are time-like.

\begin{figure}
\hspace*{0.5cm}
\psfig{file=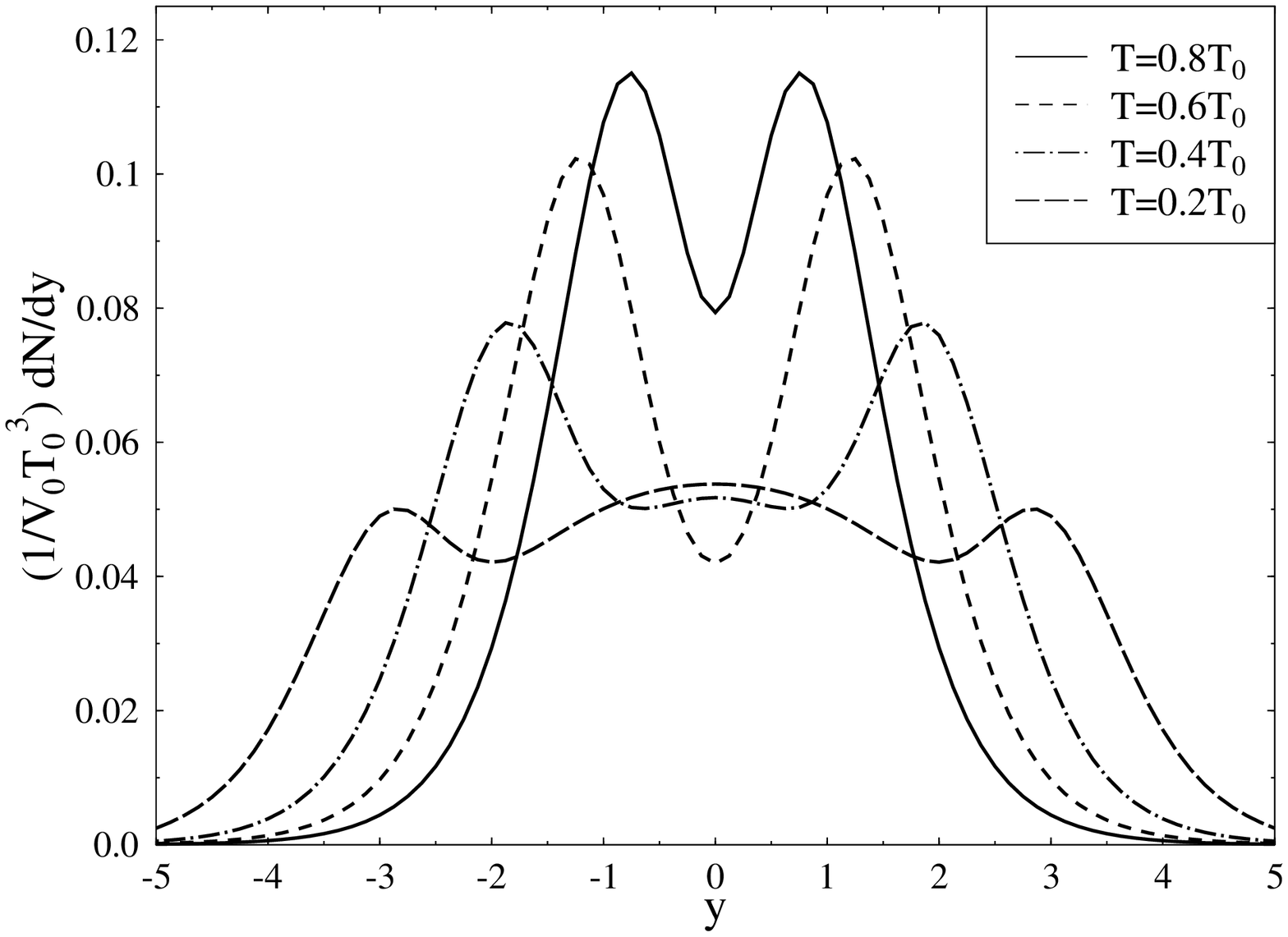,width=2.7in,height=3.8in,angle=-90}
\caption{Rapidity distributions for freeze-out along isotherms
with $T_{\rm fo}= 0.8\, T_0$ (solid), $0.6\, T_0$ (dashed), 
$0.4\, T_0$ (dash-dotted), and $0.2\, T_0$ (long dashed) in the Landau model
with a $p = \epsilon/3$ equation of state.}
\label{ris:Fig20}
\end{figure}

One suggestion to circumvent this problem was to compute the final
spectra only from contribution of particles which cross the space-like 
parts of $\Sigma$. Of course, as can be seen by comparing the
dash-dotted with the solid line in Fig.\ \ref{ris:Fig18}, 
the final spectra are dramatically different. Moreover, by neglecting
particles crossing the time-like parts, the absolute number of frozen-out
particles will also differ in the two cases. Note that the $\D N/\D y$
distribution for particles from the space-like parts of the decoupling
isotherm has a Gaussian shape in the Landau model. This was already
pointed out in Landau's original paper \cite{ris:landau} and has since
survived as the generic (but wrong) statement that Landau's model
gives rise to Gaussian rapidity distributions. In fact, there is
{\em no\/} decoupling temperature where the full rapidity
distribution including particles from the time-like parts
resembles a Gaussian, cf.\ Fig.\ \ref{ris:Fig20}.

\begin{figure}
\hspace*{2cm}
\psfig{file=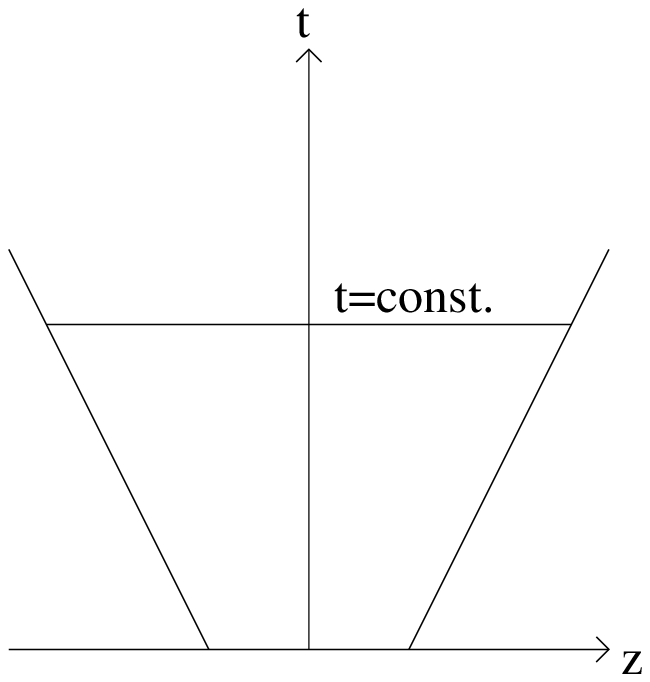}
\caption{A curve of constant time in the center-of-mass frame as
freeze-out isotherm.}
\label{ris:Fig21}
\end{figure}

\begin{figure}
\hspace*{0.5cm}
\psfig{file=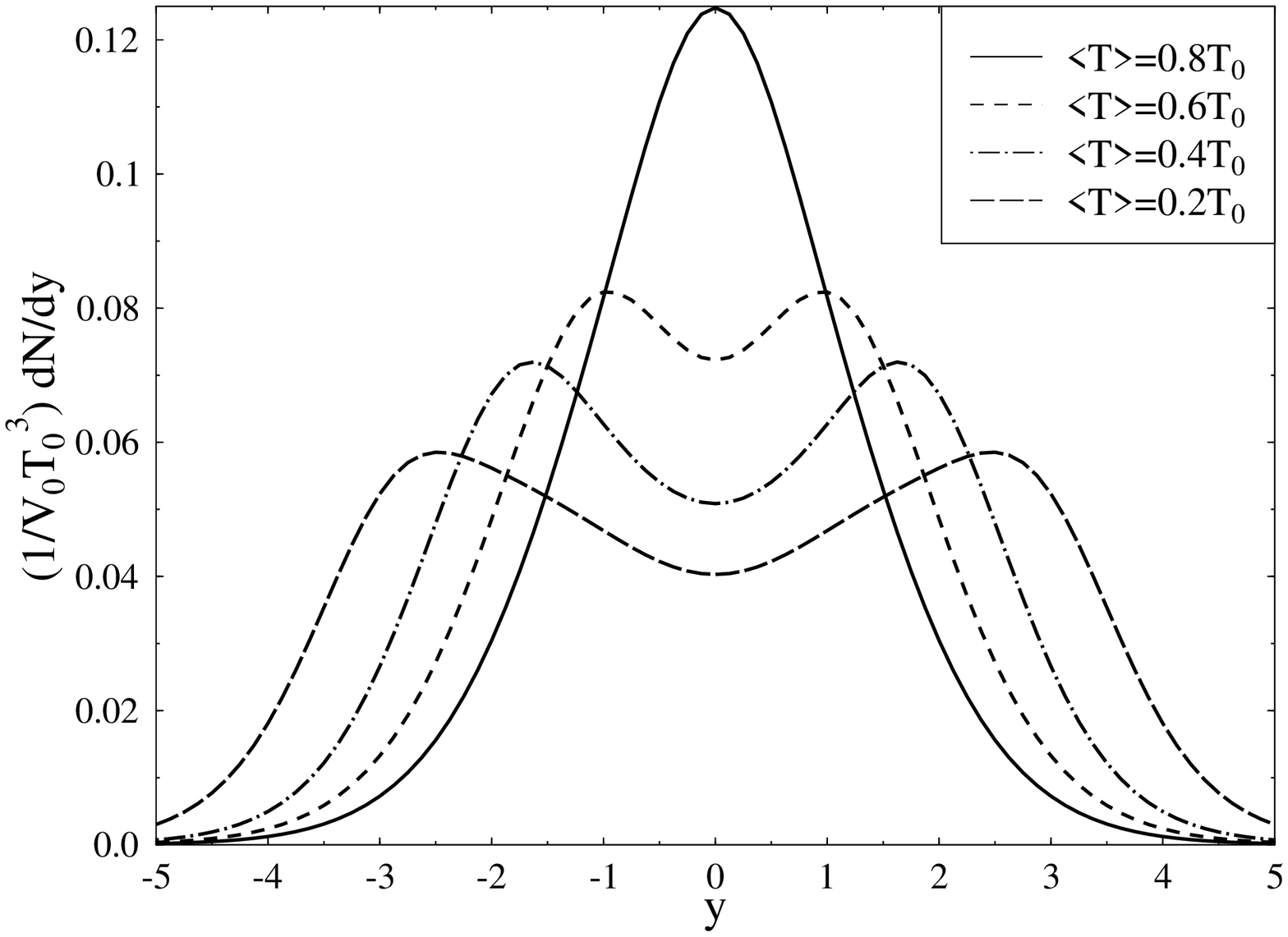,width=2.7in,height=3.8in,angle=-90}
\caption{The rapidity distribution for freeze-out along 
curves of constant time in the center-of-mass frame
defined by requiring the average temperature to be $\langle T \rangle
= 0.8\, T_0$ (solid), $0.6\, T_0$ (dashed), 
$0.4\, T_0$ (dash-dotted), and $0.2\, T_0$ (long dashed) in the Landau model
with a $p = \epsilon/3$ equation of state.}
\label{ris:Fig22}
\end{figure}

Another suggestion to circumvent the problem of negative particle numbers 
is, instead of freezing out along an isotherm which has
time-like parts, to freeze out along a surface which is space-like everywhere,
for instance, a curve of constant time in the center-of-mass frame,
cf.\ Fig.\ \ref{ris:Fig21}. In this case, all particles are accounted for,
since the decoupling surface is bounded by the lightcone, and no particle
can escape through the lightcone. The problem is, that also in this case,
the spectra differ considerably from a freeze-out at constant temperature,
cf.\ Fig.\ \ref{ris:Fig22}. This uncertainty is clearly unwanted when one
wants to quantitatively compare fluid-dynamical model predictions
with experimental data.

The correct formula to compute the number of particles which 
physically decouple from the system was given in \cite{ris:kyrill}:
\begin{equation} \label{ris:modcooper}
E \frac{\D N}{\D^3 \vec{k}} = \int_\Sigma 
\D \Sigma \cdot k \, f_0(k,x) \, \Theta(\D \Sigma \cdot k) \enspace .
\end{equation}
The additional $\Theta$-function ensures that negative contributions
to the Cooper--Frye formula are cut off. The problem with this formula
is that these negative contributions were necessary to globally conserve
energy, momentum and net charge number, cf.\ the derivation of the
conservation equations in Section 2. The violation of the conservation
equations introduced by the freeze-out prescription (\ref{ris:modcooper})
can, however, be circumvented by adjusting temperature, chemical
potential, and the average particle 4-velocity in the single-particle
distribution function $f_0(k,x)$ in (\ref{ris:modcooper}) in such a
way as to preserve the conservation laws. In other words, one
must not use temperature, chemical
potential, and fluid 4-velocity on the fluid side of the freeze-out
surface in (\ref{ris:modcooper}), but modified values
which ensure that energy, momentum, and net charge is conserved.
One way to achieve this is to assume
that the freeze-out surface actually is a conventional fluid-dynamical
discontinuity across which energy, momentum, and net charge number
are conserved. Solving the corresponding 
algebraic conservation equations (with energy-momentum tensor
and net charge current on the post freeze-out side of the discontinuity
constructed from (\ref{ris:Nmukin},\ref{ris:Tmunukin}) 
with $f_0(k,x)$ replaced by
$f_0(k,x)\, \Theta(\D \Sigma \cdot k)$) yields the required modified
values for temperature, chemical potential, and average particle
4-velocity on the post freeze-out side. For more details,
see \cite{ris:kyrill,ris:laszlo}. However, it still remains to be
shown with an explicit calculation whether this suggestion to solve
the freeze-out problem is viable in the general case.

%
%


\begin{thebibliography}{3}
\bibitem{ris:kaempfer}
To my knowledge, this treatment of the transformation problem
is due to B.\ K\"ampfer.
\end{thebibliography}
\end{document}